\documentclass[11pt, a4paper, twoside]{article}

\usepackage[utf8]{inputenc}
\usepackage[T1]{fontenc}
\usepackage{lmodern}

 \usepackage{picture}
\usepackage{stmaryrd} 
\usepackage{graphics}
\usepackage{graphicx}
\usepackage{booktabs}
\usepackage{multirow}
\usepackage{array}
\usepackage{amsmath}
\usepackage{mathtools}
\usepackage{verbatim}
\usepackage{relsize}
\usepackage[dvipsnames]{xcolor}
\usepackage{array}
\usepackage{bm}
\usepackage{hyperref}
\usepackage{todonotes}
\usepackage{setspace} 
\usepackage{lineno}
\usepackage{upgreek}
\usepackage{cases}

\makeatletter
\newcommand{\thickhline}{%
    \noalign {\ifnum 0=`}\fi \hrule height 1pt
    \futurelet \reserved@a \@xhline
}

\usepackage[width = 18cm, height = 23cm]{geometry}
\graphicspath{{figures/}{figures/discrete/}{figures/remesh_process/}}

\newcommand{\footremember}[2]{%
	\footnote{#2}
	\newcounter{#1}
	\setcounter{#1}{\value{footnote}}%
}
\newcommand{\footrecall}[1]{%
	\footnotemark[\value{#1}]%
} 

\usepackage[numbers]{natbib}

\newcommand{\seff}{\ensuremath{s_{\mathrm{eff}}}}
\newcommand{\eeff}{\ensuremath{e_{\mathrm{eff}}}}
\newcommand{\feff}{\ensuremath{f_{\mathrm{eff}}}}
\newcommand{\ft}{\ensuremath{f_{\mathrm{t}}}}
\newcommand{\Gt}{\ensuremath{G_{\mathrm{t}}}}
\newcommand{\rf}{\ensuremath{r_{\mathrm{f}}}}
\newcommand{\rt}{\ensuremath{r_{\mathrm{t}}}}
\begin{document}

\title{Adaptive discretization refinement for discrete models of coupled mechanics and mass transport in concrete}
\author{Jan Ma\v{s}ek\footremember{Brno}{Brno University of Technology, Faculty of Civil Engineering, Brno, Czechia}\footremember{email}{Corresponding author: jan.masek1@vut.cz } \and Josef Kv\v{e}to\v{n}\footrecall{Brno} \and Jan Eli\'{a}\v{s}\footrecall{Brno}}
\date{}

\maketitle 

\section*{Abstract}
An adaptive discretization refinement strategy for steady state discrete mesoscale models of coupled mechanics and mass transport in concrete is presented. Coupling is provided by two phenomena: the Biot's theory of poromechanics and an~effect of cracks on material permeability coefficient. The model kinematics is derived from rigid body motion of Voronoi cells obtained by tessellation of the domain. Starting with a~coarse discretization, the density of Voronoi generator points is adaptively increased on the fly in regions where the maximum principal stress exceeds a~chosen threshold. Purely elastic behavior is assumed in the coarse discretization, therefore no transfer of history/state variables is needed. Examples showing (i) computational time savings achieved via the adaptive technique and (ii) an~agreement of the outputs from the fine and adaptive models during simulations of hydraulic fracturing and three-point bending combined with a~fluid pressure loading are presented.

\section*{Keywords}
adaptive discretization; concrete; coupling; discrete model; mechanics; fracture; mass transport; simulation; hydraulic fracturing

\section{Introduction}

Mesoscale mechanical models with kinematics derived from the rigid body motion (as proposed by~\citet{Kaw78}) are one of the most reliable and robust way to simulate pre- and post-critical mechanical behavior of concrete. These high-fidelity models represent concrete as an~assembly of rigid bodies interconnected by cohesive contacts. Thanks to discrete jumps in the displacement field, these models are ideal for simulating fracture. The vectorial constitutive formulation at the contact facets is simpler than the traditional tensorial form and it automatically provides the orientation of cracks. Another major advantage is the occurrence of stress oscillations due to material heterogeneity. These oscillations are crucial under compressive loading  because they result in splitting cracks parallel to the loading direction~\citep{WanLuz-22,CusBaz-03II,CusMen-11}. 

There is a~large number of different discrete models for concrete mechanical behavior developed in the literature, see e.g. the  recent review paper by~\citet{BolEli-21}. Some of the models are built at macroscale as a~discretization technique of homogeneous continuum~\citep{BolSai98,HwaBol-20,HwaBol-20b,KhmSte21,DawSho-22}. They benefit from possibility of freely selecting size of the discrete units since the discretization is \emph{not physical}, but they suffer from necessity to regularize energy dissipation accordingly~\citep{BerBol06}.  Similarly, there are models with \emph{non-physical} discretization with elements smaller than size of material heterogeneities~\citep{GraGre-12,LukSav-16,LuoAsa-22,YadYid-22}, where the regularization of dissipated energy is more or less automatically provided by a~projected mesostructure. These detailed models are, however, computationally expensive and can be applied only to material volumes of limited size. Moreover, all models with \emph{non-physical} discretization should ideally eliminate stress oscillations caused by the irregularity of the domain tessellation.

It seems to the authors that the most appealing are discrete models of concrete with \emph{physical} discretization, where each rigid body encompasses one larger mineral aggregate and the surrounding matrix~\citep{CusBaz-03I,EliVor20,EliVor-15,MonCif-17,AvaPun-21}. Arguably the most prominent representant of this group is the Lattice-Discrete Particle Model (LDPM) developed by Gianluca Cusatis and co-workers~\citep{CusMen-08,CusPel-11,ZhoMad-22} but implemented also by other research groups~\citep{FasBol-18,FasIch-22,LifFad-22}.  Mechanical response, stress fluctuations or crack openings do not need to be independent on the size of the rigid bodies; their size is dictated by the size of material heterogeneities. Thanks to a low number of kinematic parameters (only 6 degrees of freedom for one rigid body in three dimensions), these mesoscale models can be directly used for analysis of small concrete structural components. For example,  \citet{BhaSha-21} or \citet{KhoMat-19} employed discrete models with \emph{physical} discretization to study cracking in bent reinforced beams,  \citet{MarVor-17} analyzed adhesive anchors, \citet{AvaPun-22} studied cracking due to corrosion of reinforcing steel. 

\begin{figure}[b!]
\centering{\includegraphics[width=10cm]{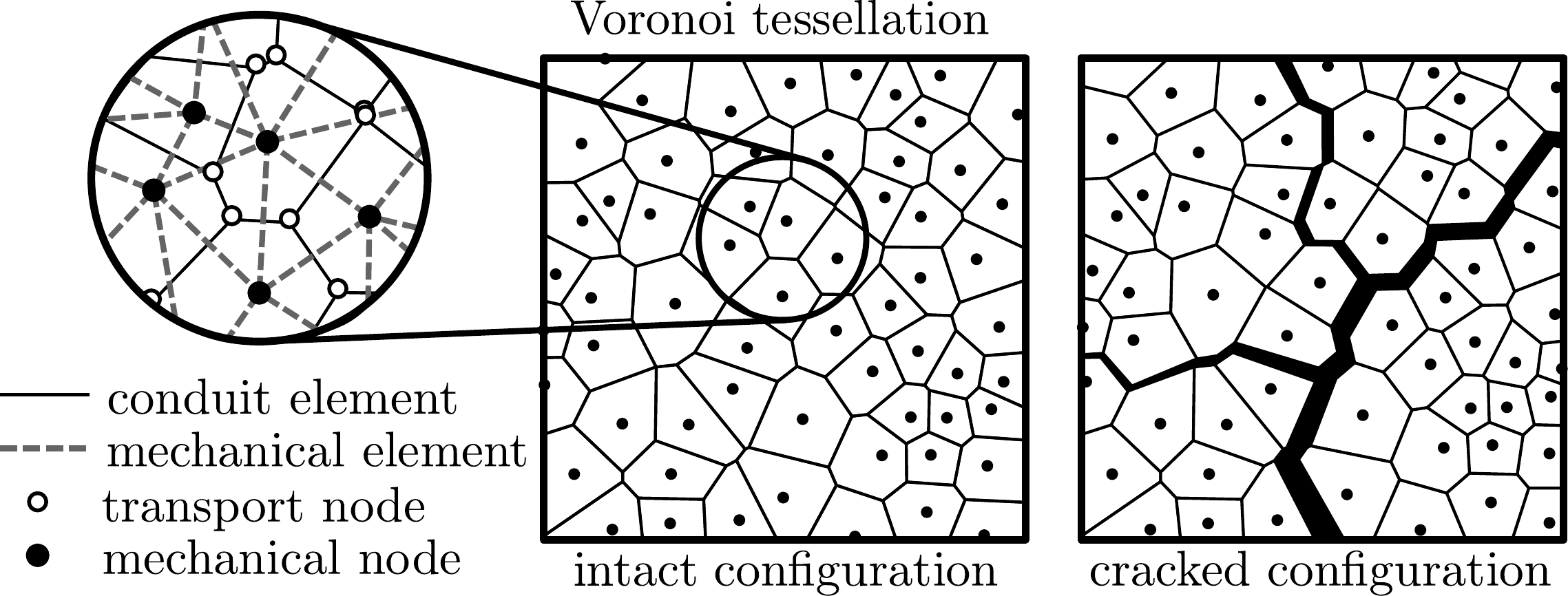}
\caption{Left-hand side: a tessellation of a~2D domain into (i) ideally rigid Voronoi cells for the mechanical problem and (ii) Delaunay simplices representing control volumes for the mass transport problem; right-hand side: deformed state with opened cracks creating channels for mass transport.}
\label{fig:tes}}
\end{figure}

The detailed representation of a complex, interconnected and highly anisotropic crack structure in the material makes the discrete models ideal to be employed in the coupled multi-physical analysis of mechanics and mass transport. Such a coupled analysis is crucial for assessment of structural durability. The initial coupling technique in the discrete framework by ~\citet{BolBer04} proposed to use identical nodes (the Voronoi generator points) to define mechanical and pressure fields. Location of the pressure nodes was later updated by adding them also to the centers of Voronoi edges~\citep{NakSri-06}; then the whole mass transport problem was transferred to a~dual network~\citep{Gra09,GraBol16}, where the pressure nodes are associated with Voronoi vertices, see Fig.~\ref{fig:tes}. The latter technique is advantageous because the conduit elements are perpendicular to the mechanical contacts where cracks develop. They can easily account for presence of oriented cracks and therefore correctly evaluate an anisotropic material permeability tensor in the cracked domain. A~minor disadvantage of the latter approach is an~increased computational burden since the number of Voronoi vertices is much larger than number of Voronoi generator points in three dimensions. 

Many publications use discrete models with combination of primary and dual network for coupled multi-physical analysis. The coupling only through crack opening is often referred to as a~\emph{one-way} coupling because the mechanical problem is independent on the transport one~\citep{Gra09,FahWhe-17,AthWhe-18,SinSav-22}. Another important coupling source is the poromechanical Biot's theory, where the pressure field interacts with the rate of volumetric deformation and affects tractions at the contacts. Examples of such \emph{two way} coupled discrete models can be found in Refs.~\citep{LiZho-18,SheMon-21,LefNou-20,GraJir-19}. 

Despite the simplified kinematic description, the coupled mesoscale discrete models with \emph{physical} discretization are still relatively computationally expensive and cannot be used for larger volumes of material. There are several options for reducing the computational burden such as reduced order modeling~\citep{KerPas-12,CecZho-18}, coarse graining~\citep{LalRez-18}, quasicontinuum~\citep{TadOrt-96,SorBis14,RokPee-17,MikJir17} or homogenization~\citep{RezCus16,EliYin-22,EliCus22}. 

Another option for reducing the computational cost is an adaptive discretization. Traditionally, the adaptive technique has been explored in finite element models to locally improve quality of the approximated displacement field~\citep{BabRhe78,ZieZhu87,PatJir04}. The change of mesh involves a transfer of history variables onto the newly created integration points. In the discrete framework, one should rather speak about replacing locally the cheap, low-fidelity model with a~detailed and expensive high-fidelity model in areas where needed. The low-fidelity model is typically the finite element approximation of a homogeneous continuum~\citep{BolShi-96,CorMat-20}. However, the connection between finite elements and discrete structure is problematic because of the random stress oscillations in the discrete structure. Removing these oscillations opens a~possibility of a~smooth connection with the finite elements~\citep{HwaJi-22,HwaBol-22}, but it takes away several advantages of mesoscale discrete models. For this reason, we develop the adaptive discretization technique completely within the discrete framework; the low fidelity model is also a~discrete model, but with a~\emph{non-physical} macroscopic discretization. 

Following the work of~\citet{Eli16} where the  adaptive refinement from coarse to mesoscale discrete model was introduced for mechanics, we extend the concept for \emph{two-way} coupled systems of mechanics and mass transport in saturated heterogeneous solids. Adaptive technique developed in Ref.~\citep{Eli16} relied on the independence of the linear steady state mechanical solution on the size of the rigid bodies. The solution of the scalar potential field (in our case pressure) in linear regime is also independent on the size of the rigid bodies, see for example Ref.~\citep{AsaKun-14}. Therefore, one can use  a~coarse \emph{non-physical} representation of the material as long as it stays linear. Whenever the material state approaches the nonlinear regime, the refinement takes place; the low-fidelity coarse model is locally replaced by the high-fidelity mesoscale model with \emph{physical} discretization. Since the coarse model is linear, there are no history variables to transfer. Also the connection of the two discrete models is straightforward since both of them are formulated using the same equations. 

The adaptive technique for two-way coupled discrete models is verified by numerical simulations on concrete specimens comparing loading forces, fluxes, crack patterns and computational times of (i) the coarse, low-fidelity model, (ii) the computationally demanding high-fidelity mesoscale model and (iii) the efficient model with the proposed adaptive refinement.
Specifically, we first present coupled numerical simulations of 3D concrete specimens loaded  by a~combination of three-point-bending and fluid pressure at the bottom face. Then, a simplified 2D model of concrete fracture due to reinforcement bar corrosion in two different setups is presented. 

All the models used in the paper are implemented in an~open-source in-house computer code named OAS. 

\section{Discrete mesoscale model} 
\label{sec:discrete_model}

A heterogeneous material is represented by a~system of interconnected, ideally rigid particles. These polyhedral particles are assumed to represent larger mineral grains in concrete as well as surrounding matrix. Particle geometry is produced by the Voronoi tessellation built on~a set of generator points randomly distributed within~a volume domain~\citep{BolSai98}, Fig.~\ref{fig:tes}. The generator points are sequentially added into the domain with a~given minimum distance $l_{\min}$, which dictates the size of the material heterogeneities.  

There are more sophisticated ways of generating the mesoscale internal structure of concrete. For example by a~random placement of non-overlaping spherical inclusion distributed according to a~given sieve curve. One can then employ the power/Laguerre tessellation~\citep{Eli17,EliCus22} or some tailored tessellation algorithm~\citep{CusPel-11}. These tessellation techniques are not used in the present paper but there is, in principle, no restriction in combining them with the adaptive refinement developed here.

The Voronoi generator points, called mechanical nodes and denoted $I$ and $J$ hereinafter, bear degrees of freedom (DoF) of the mechanical problem. In two dimensions, there are two translations, $\mathbf{u}$, and one rotation, $\theta$; in three dimensions three translations, $\mathbf{u}$, and three rotations, $\bm{\uptheta}$, are present. The Voronoi vertices, called transport nodes  and denoted $P$ and $Q$ hereinafter, have one associated DoF defining pressure, $p$.   

\subsection{Kinematic equation}
\begin{figure}[tb!]
\centering{\includegraphics[width=15cm]{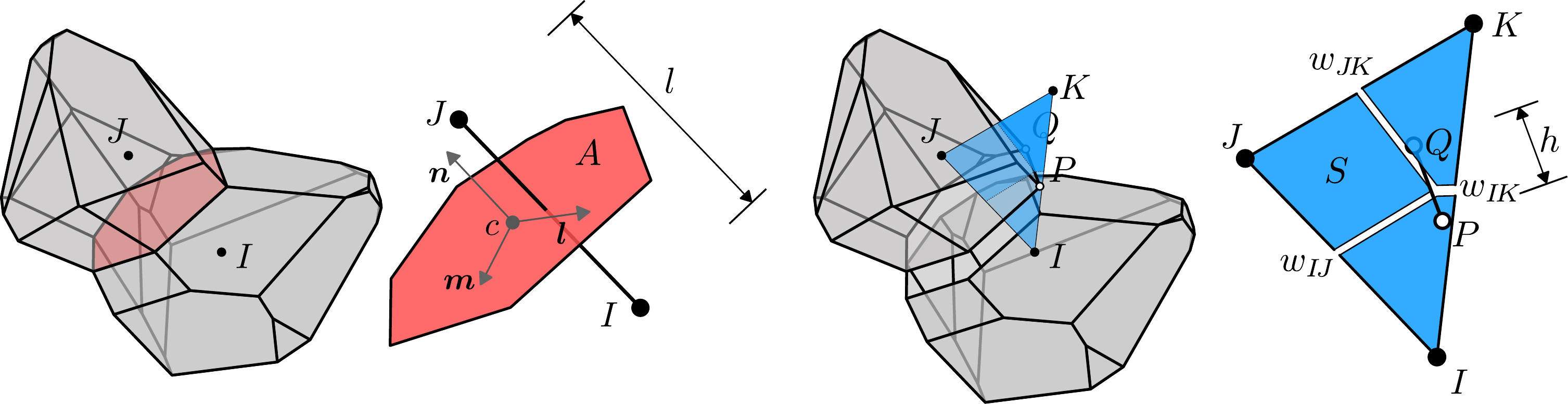}
\caption{Left-hand side: two rigid bodies in contact via a~mechanical element; right-hand side: a~conduit element formed by a~contact of two simplices associated with transport nodes $P$ and $Q$ and a simplified representation of crack openings in adjacent mechanical elements. }
\label{fig:facet}}
\end{figure}

Mechanical elements connect two mechanical nodes; they run along the Delaunay edges. Their length is denoted $l$, contact area is $A$ and the centroid of the contact face is $\mathbf{c}$. One of such elements is depicted in Fig.~\ref{fig:facet} on the left-hand side. Faces of the mechanical elements connecting one node with the neighbors always create a convex polyhedron -- a Voronoi cell.  The chosen tessellation scheme ensures perpendicularity between the contact face and the line connecting $I$ and $J$. The normal direction $\mathbf{n}$ therefore describes also the direction of that line. There are also tangential directions $\mathbf{m}$ and $\mathbf{l}$ (in 2D only one) chosen 
randomly  to form an~orthonormal basis with $\mathbf{n}$.

The conduit elements run along Voronoi edges. The length of the conduit element is denoted $h$, its area is $S$, see Fig.~\ref{fig:facet} on the right-hand side. The faces of the conduit elements create Delaunay triangles (2D) or tetrahedrons (3D), respectively. The Voronoi tessellation ensures the perpendicularity between the conduit element face and the line connecting nodes $P$ and $Q$.   

The strain vector, $\mathbf{e}$, and the pressure gradient scalar, $g$, are given by the kinematic equation. The pressure gradient is simply the pressure difference divided by the length of the conduit element. The strain is computed as a~displacement jump between the rigid bodies at the integration point, $\mathbf{c}$, divided by the length of the mechanical elements and rotated into the local reference system
\begin{align}
g &= \frac{p_J-p_I}{h} \\   
\mathbf{e} &= \frac{1}{l}\llbracket\bm{u}\rrbracket \cdot \bm{\uprho}
\end{align}
where $\bm{\uprho}$ is the~rotation tensor of directional cosines of the axes $\mathbf{n}$, $\mathbf{m}$ and $\mathbf{l}$ of the local reference system in columns ($\bm{\uprho}=\left(\mathbf{n}\ \mathbf{m}\ \mathbf{l} \right)$ in 3D and $\bm{\uprho}=\left(\mathbf{n}\ \mathbf{m}\right)$ in 2D, respectively). Only one integration point $\mathbf{c}$ located at the facet centroid is used.

The displacement jump, $\llbracket\bm{u}\rrbracket$,  evaluated in the global reference system from rigid body kinematics under the assumption of small rotations read
\begin{align}
\llbracket\bm{u}\rrbracket = \begin{cases} \mathbf{u}_J - \mathbf{u}_I + \bm{\mathcal{E}}:\left(\bm{\uptheta}_J \otimes \mathbf{c}_J- \bm{\uptheta}_I \otimes \mathbf{c}_I\right) & \mbox{in\ 3D}\\
\mathbf{u}_J - \mathbf{u}_I - \bm{\mathcal{E}}\cdot\left(\uptheta_J \mathbf{c}_J-\uptheta_I \mathbf{c}_I\right) & \mbox{in\ 2D} \end{cases}
\label{eq:displ_jump_3D}
\end{align}
where $\bm{\mathcal{E}}$ is the antisymmetric Levi-Civita permutation tensor. 

There is also a~rotational jump between the rigid bodies which would give the curvature at the contact. The constitutive formulation does not assume any rotational stiffness of the inter-particle contacts, therefore the curvature and couple traction is omitted here.

\subsection{Constitutive equation}
The \emph{mechanical constitutive equation} provides traction in the solid, $\mathbf{s}$, dependent on the strain vector, $\mathbf{e}$. It is formulated in the local coordinate system with directions denoted by indices
$N$ (normal) and $M$ and $L$ (tangential, $L$ present only in 3D). The formulation is based on the Confinement-Shear Lattice (CSL) model introduced by \citet{CusCed07}, however the confinement effect is neglected and non-linearity is driven by non-decreasing damage. The number of material parameters is reduced to 4: normal elastic constant $E_0$, tangential/normal stiffness ratio $\alpha$, tensile strength $\ft$ and tensile fracture energy $\Gt$. Remaining CSL material parameters are computed using recommendations from Ref.~\citep{CusCed07}. The same constitutive model was used to analyze effects of randomness on mechanical behavior~\citep{EliVor-15,EliVor20} or, more importantly, to the developed adaptive refinement scheme for purely mechanical discrete model~\citep{Eli16}.

Tractions in the solid in the normal and tangential directions read
\begin{align}  \label{eq:solid_stress}
s_N &=  (1-d) E_0 e_N & s_M &=  (1-d) E_0 \alpha e_M & s_L &=  (1-d) E_0 \alpha e_L 
\end{align} 
where the non-decreasing damage parameter $d$ ranges from 0 (healthy material) to 1 (completely damaged material).

The damage parameter depends on the effective traction, $\seff$, and the effective strain, $\eeff$, $d \geq 1-\seff/E_0\eeff$. The effective strain is defined as $\eeff=\sqrt{e_N^2+\alpha e_T^2}$, $e_T= \sqrt{e_M^2+e_L^2}$ in 3D and $e_T= e_M$ in 2D. 
The effective stress is $\seff =  \feff \exp\left( K/\feff \left\langle \chi -\feff/E_0\right\rangle\right)$, the angled brackets return the positive part, $\feff$ is the effective strength, $K$ is the initial slope in the inelastic regime and $\chi$ represents the loading history.
\begin{align}
\feff &= \begin{cases} \frac{16\ft}{\sqrt{\sin^2\omega+\alpha\cos^2\omega}} & \omega<\omega_0 \\ \ft \frac{4.52 \sin\omega-\sqrt{20.0704 \sin^2\omega+9 \alpha\cos^2\omega }}{0.04\sin^2\omega-\alpha \cos^2\omega} & \omega\geq\omega_0 \end{cases} \label{eq:feq}
\\
K &= \begin{cases} 0.26E_0\left[ 1-\left( \frac{\omega+\pi/2}{\omega_0+\pi/2}\right)^2\right] & \omega<\omega_0 \\ -K_{\mathrm{t}}\left[ 1-\left( \frac{\omega-\pi/2}{\omega_0-\pi/2}\right)^{n_t}\right] & \omega\geq\omega_0  \label{eq:K}
\end{cases}
\\
\chi &= \begin{cases} \eeff & \omega<\omega_0 \\ \eeff\frac{\omega}{\omega_0} + \sqrt{\max e_N^2 + \alpha\max e_T^2}\left(1-\frac{\omega}{\omega_0}\right) & \omega_0\leq\omega<0 \\ \sqrt{\max e_N^2 + \alpha\max e_T^2} & \omega\geq 0
\end{cases}
\end{align} 
The variables $\max e_N$ and $\max e_T$ are maximum normal and tangential strains experienced by a~contact during the simulation. The direction of straining, $\omega$, is given by the ratio of the normal and tangential strains, $\tan \omega=e_N / \sqrt{\alpha} e_T$. The transitional straining direction, $\omega_0$, is the direction at which the first branch of Eq.~\eqref{eq:feq} equals to the second branch.

The initial strain softening slope, $K$, is defined using the slopes for pure tension ($K_{\mathrm{t}}=2E_0\ft^2 l/(2E_0\Gt - \ft^2l)$) and pure shear ($K_{\mathrm{s}}=18\alpha E_0\ft^2 l/(32\alpha E_0\Gt - 9\ft^2l)$). The power $n_t$ from Eq.~\eqref{eq:K} is governing the transition in the tension-shear regime and reads $n_t = \ln\left(K_{\mathrm{t}}/(K_{\mathrm{t}}-K_{\mathrm{s}})\right)/\ln\left( 1-2\omega_0/\pi\right)$.

The total tractions, $\mathbf{t}$, acting at the facets include also the coupling with the pressure field according to the Biot's theory. The total normal traction is decreased by the pressure multiplied by the Biot's coefficient, $b$, ranging from 0 to 1. 
\begin{align}  
t_N &=  s_N - bp & t_M &= s_M & t_L &=  s_L \label{eq:total_traction}
\end{align} 

Note that lumping the elastic material behavior into a~normal and shear contact between two nodes limits the macroscopic Poisson's ratio of the assembly~\citep{CusPel-11,Eli20}. Despite a number of successful attempts to remedy this limitation~\citep{AsaIto-15,AsaAoy-17,CusRez-17,CelLat-17,RojZub-18}, the Poisson's ratio of discrete models is still not fully resolved and understood.

The \emph{mass transport constitutive equation}, $j=\lambda g$, relates the flux, $j$, and the pressure gradient, $g$. It features the permeability coefficient, $\lambda$, dependent on the normal crack opening, $w_N=e_N ld$, computed in the mechanical part of the model. The total permeability coefficient is expressed as a~summation of the permeability coefficient of the intact material and an~effect of cracks according to Ref.~\citep{GraBol16} for 3D and Ref.~\citep{FahWhe-17} for 2D model, respectively 
\begin{align}
\lambda = \begin{cases}
\dfrac{\rho\kappa}{\mu} + \dfrac{\xi\rho}{12\mu S}\sum_{i=1}^3 w_{Ni}^3 l_{\mathrm{c}i} &\mbox{in 3D}\\
\dfrac{\rho\kappa}{\mu} + \dfrac{\xi\rho w_{N}^3}{12\mu S}  &\mbox{in 2D}
\end{cases} \label{eq:lambda}
\end{align}
where $\rho$ is the density of the fluid, $\kappa$ is the material permeability, $\mu$ is the viscosity, $\xi$ is the crack tortuosity parameter and $l_{\mathrm{c}}$ is the~crack length of associated mechanical elements, see Refs.~\citep{Gra09,GraBol16,FahWhe-17} for details.

The two way coupling between the mechanical and transport part is provided by (i) Eq.~\eqref{eq:lambda}, where mechanical crack openings modify the material permeability tensor and (ii) Eq.~\eqref{eq:total_traction}, where the pressure affects the total normal traction at the contacts between mechanical particles.

\subsection{Balance equation} \label{sec:balance_eq}
Three sets of balance equations need to be satisfied: the balances of linear and angular momentum and fluid mass balance. The linear and angular momentum balances are formulated separately for each rigid body $I$, while the mass balance is satisfied for each simplex $P$. We limit ourselves to the steady state behavior, for which the linear and angular momentum balance in the global reference system for both 3D and 2D (only $z$ becomes scalar) space reads
\begin{align}
 \sum_J A~\mathbf{t} \cdot \bm{\uprho}&= -  V \mathbf{b}  \label{eq:linear_balance} \\
\sum_J A~\bm{\mathcal{E}}:(\mathbf{c}_I\otimes \mathbf{t})&= - V \mathbf{z} - V \bm{\mathcal{E}}:(\bar{\mathbf{x}}_I\otimes \mathbf{b})\label{eq:angular_balance}
\end{align}
Variables $\mathbf{b}$ and $\mathbf{z}$ represent volume loads acting within the rigid body and $\bar{\mathbf{x}}_I$ is the centroid of the rigid body $I$. Note that the second equation sums on the left-hand side only moments of tractions, there are no couple tractions considered. The mass balance sums outward fluxes $j$ towards all neighbors $Q$ and involves also the source or sink term $q$.
\begin{align}
\sum_{Q} S j  = Wq \label{eq:mass_balance}
\end{align}
The coupling due to the volumetric deformation rate as developed in Refs.~\citep{LiZho-18,EliCus22} is not included because of the steady state nature of the equation.

\section{Adaptive refinement scheme}
The high-fidelity mesoscale model introduced in the previous section relies on the  correspondence between discrete bodies and mineral aggregates in concrete. The inelastic behavior, for which such a model is needed, develops often only locally in a~relatively small portion of the whole specimen. Consequently, a~large computational effort is consumed unreasonably by calculation of elastic material response, which can be easily solved by a coarser model with a \emph{non-physical} discretization.
It is therefore convenient to use such a~fine discretization only in regions where it is needed, i.e., where inelastic processes take place. 
Such a~mixed discretization can advantageously be generated in analyses where the regions of inelastic behavior are known in advance. In the remaining cases where the location of the inelastic zone is unknown at the beginning of the simulation, the adaptive refinement scheme helps to maintain the quality of the results while keeping the computational cost at a~reasonable level at the same time. 

The present article addresses an adaptive refinement technique for the coupled mass transport-mechanical discrete model as described in Sec.~\ref{sec:discrete_model}.
The geometrical representations of mechanical and conduit elements are interconnected by duality of Voronoi--Delaunay structures. It is therefore enough to refine the Voronoi generator points to achieve a simultaneous refinement in both structures. Since the only non-linearity in the mass transport part comes from cracking, the adaptive refinement is governed solely by the mechanical part of the model.

The adaptive refinement is initiated whenever a~chosen criteria in any rigid particle exceeds a~given threshold.
Then, a neighborhood of such a~particle is refined.
The refinement criteria and size of the refined neighborhood are model input parameters. In principle, any refinement criteria is suitable. If chosen poorly, there might be large discrepancies between original elastic and refined inelastic states. Various criteria for concrete failure can be adopted from literature~\cite{Ott77,BozChe87}. In the presented work focused on tensile failure, the Rankine stress criteria is utilized. The maximum principal stress in the solid part of the model must be below a~prescribed threshold, which is chosen as $70\,\%$ of material tensile strength, $\ft$, as justified in Ref.~\cite{Eli16}.

The maximum principal stress in a particle is computed from eigen-decomposition of the tensorial stress. Such a~quantity is not directly available in the model, because the particles are rigid and interact via vectorial tractions.  However, an~average stress tensor of each particle $I$ can be evaluated according to~\citep{Web66,BarVar01} as a~direct product of the position of integration points at all contacts and neighboring particles $J$ with the traction in the solid part at that integration point
\begin{align}
\langle\bm{\upsigma}_I\rangle_{\mathrm{sym}} &= \left[\frac{1}{V}\sum\limits_{J}  \mathbf{c}_{IJ}  \otimes \mathbf{s}_{IJ}\right]_{\mathrm{sym}}
\label{eq:fabric_stress}
\end{align}
Note that the above equation does not work with the total tractions but only with the tractions of the solid part. Consequently, also the tensorial stress from Eq.~\eqref{eq:fabric_stress} neglects the effect of pressure. This allows us to identify regions where constitutive equation approaches inelastic states due to the tensile stress limit, even though the total stress might be in compression when the hydrostatic component of pressure is added. Also note that the stress must be symmetrized as the underlying continuum description corresponds to Cosserat type of medium~\citep{RezCus16,EliCus22} with non-symmetric stress tensor.

The refinement process is now described in detail. 
Initially, the domain is discretized into relatively coarse particles with size dictated by gradients of the strain field (excessively large particles might approximate the elastic stress and pressure state too poorly). The minimum distance of Voronoi generator points in the coarse structure is denoted $l_{\min,\mathrm{c}}$. Such a~model can be used only in the linear regime, because the coupling via cracking and a~inelastic mechanical behavior are both dependent on the  size of rigid particles.

After every step of the iterative solver (in the state of balanced linear and angular momentum and fluid flux) at time $t+\Delta t$, the average solid stress tensor is computed for every particle according to Eq.~\eqref{eq:fabric_stress}. The refinement criteria is evaluated -- in our case the Rankine effective stress is compared to 0.7 multiple of the tensile strength -- in all particles except those belonging to the refined \emph{physical} discretization. Critical particles approaching the inelastic regime are detected. If there are no such critical particles that exceed the threshold stress, the simulation continues to the following time step. Otherwise, the whole time step is rejected and simulation time is reset to $t$. Also, internal history variables at time $t$, namely the damage parameter, $d$, and maximum strains, $\max e_N$ and $\max e_T$, of the contacts in the refined part of the model are saved. 

\begin{figure}[tb!]
    \centering
    \def\svgwidth{\textwidth}
    \small
      \includegraphics{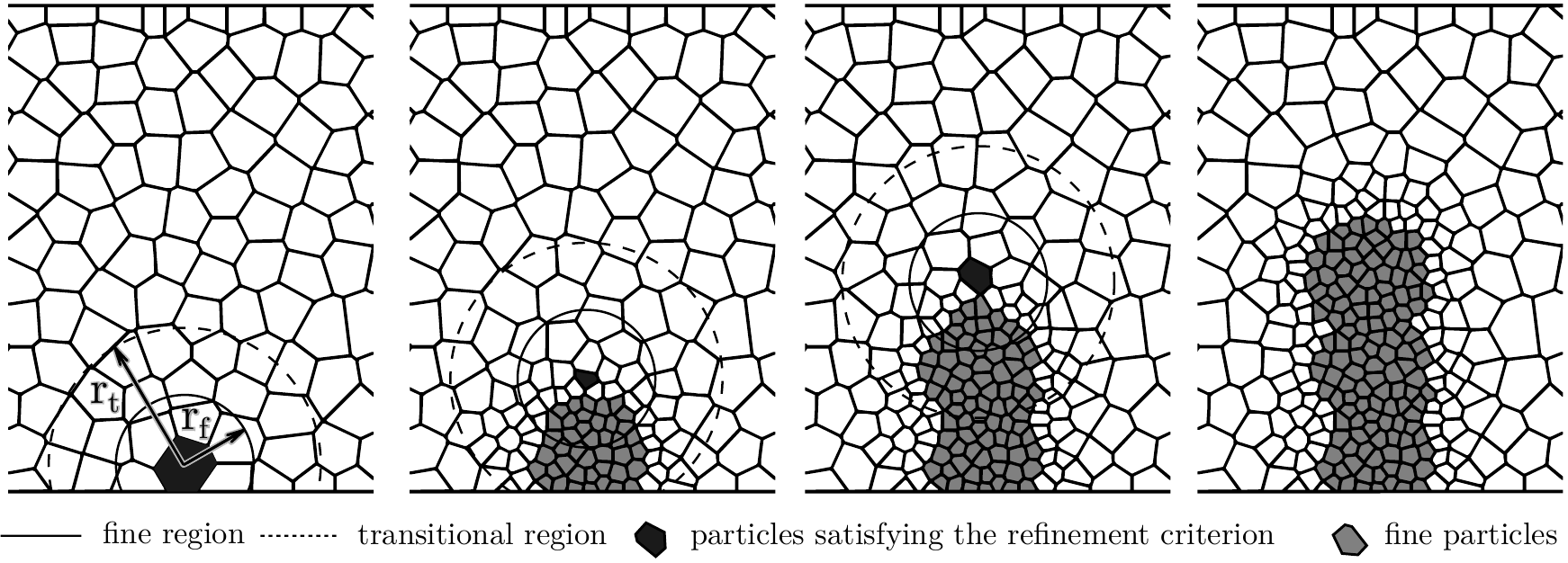}
    \caption{The process of geometry update, from left to right: a region of  a specified radius around a critical particle that satisfies the refinement criteria is removed and refined. Gray fine particles are not refined anymore, since they correspond to the material internal structure.} 
    \label{fig:remesh_scheme}
\end{figure}

Next the Voronoi generator points defining the model geometry are updated and regional remeshing to the fine \emph{physical} discretization is conducted.
The the minimum distance of Voronoi generator points in the \emph{physical} discretization is denoted as $l_{\min,\mathrm{f}}$.
{\color{black} Within the spheres of radii $\rt$ around centroids of every critical particle, all the generator points not belonging to the fine \emph{physical} discretization are removed, see Fig.\,\ref{fig:remesh_scheme}. Those from the \emph{physical} discretization  are kept.
New generator points are then inserted into the cleared regions with (i) the minimum distance corresponding to the \emph{physical} discretization, $l_{\min,\mathrm{f}}$, in the spheres of radii $\rf$ and (ii) %
with the minimum distance linearly increasing  from $l_{\min,\mathrm{f}}$ to $l_{\min,\mathrm{c}}$ as the distance from the critical particle changes from $\rt$ to $\rf$.
These two radii, $\rf$ and $\rt$, are model parameters. The model geometry is then generated again based on the updated set of generator points. 

The transitional region was introduced in \cite{Eli16} to avoid a surrounding layer of oblong particles.
The purpose of the transitional region is to mitigate possible anisotropy arising from an  abrupt change of discretization density in the model, see studies in \cite{Eli17}.
The larger the radius $\rf$ is, the fewer remeshing steps are needed but the more DoF are introduced into the model.
Extremely small $\rf$ might affect the accuracy of the computed response. \citet{Eli17} performed a study attempting to optimize these radii with respect to the total execution time, but such results are strongly problem dependent. Based on a similar study, the present work sets these radii to $\rf = 5 \, l_{\min,\mathrm{f}}$ and $\rt = 8 \, l_{\min,\mathrm{f}}$.
}

Finally, the saved history variables are loaded for the updated model. The geometrical structure of the area with \emph{physical} discretization is not changed so the contacts for which the history was saved appear in the new discretization unchanged. The iterative search for equilibrium at time $t$ is repeated before another load is imposed.
The adaptive geometry update is advantageous regarding the number of degrees of freedom. However, it should be mentioned that the update of the internal structure might consume additional computational time. The comparisons presented in the further sections show a~significant reduction of computational cost when comparing simulations with \emph{physical} discretization to those with the adaptive refinement. The actual speed-up factor achieved by the adaptive technique includes also the time needed to update the model internal structure.

\section{Verification examples}

\subsection{Three point bending combined with pressure loading}
The first verification example attempts to capture the influence of water pressure within propagating cracks in concrete. The geometry of the beam loaded by three-point bending was inspired by Ref.~\citep{gregoire2013failure}. Additionally, the beam is also loaded by a~constant water pressure on its bottom face. The water pressure eventually propagates into the cracks emerging at the bottom surface due to the mechanical loading, as studied by \citet{slowik2000water}. Depending on the value of the Biot coefficient, the pressure of water within propagating cracks further contributes to the crack opening. 

\begin{figure}[htbp]
    \centering
    \includegraphics[width=9cm]{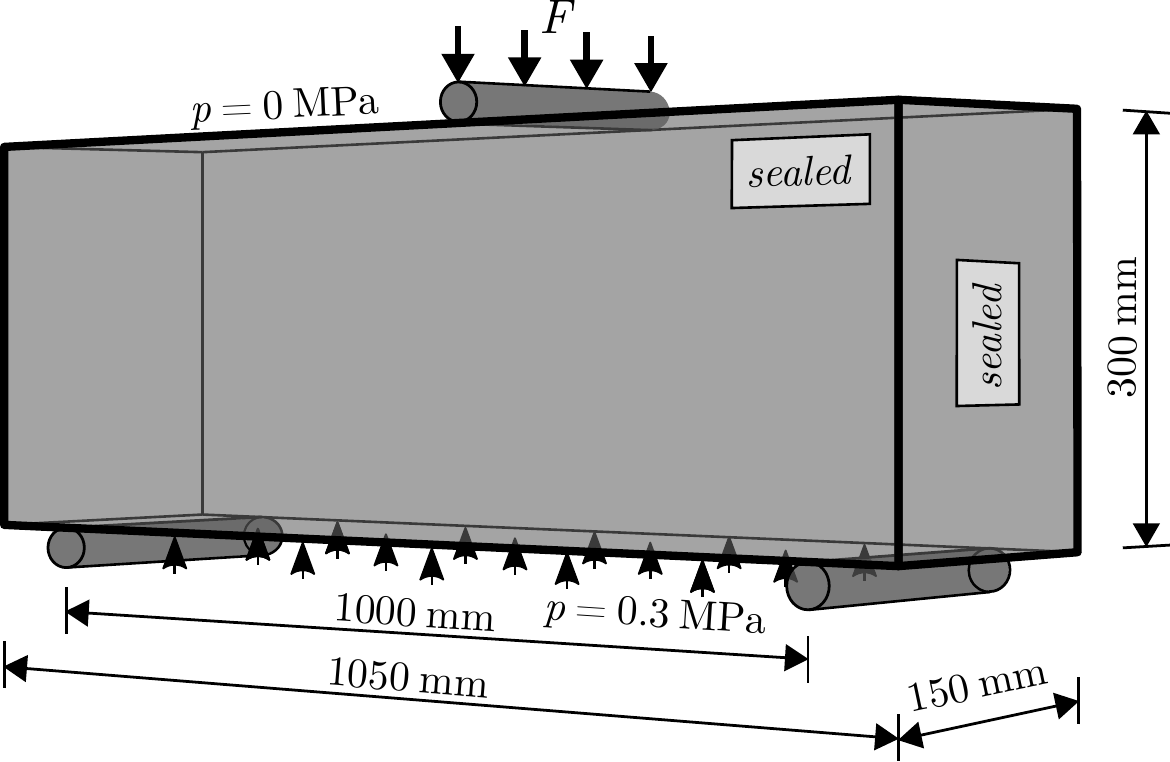}
    \caption{Coupled three-point bending experiment setup.}
    \label{fig:3pbgeometry}
\end{figure}

The numerical experiment is conducted in steady state, meaning that the duration of the process is long enough so the capacity and inertia terms can be neglected. The mechanical loading is controlled by the crack opening displacement at the bottom surface to avoid unstable post-peak crack propagation.
The geometry and loading scheme is shown in Fig.~\ref{fig:3pbgeometry}. The modeled specimen is an~unnotched beam of span 1\,m, depth 0.30\,m, and thickness 0.15\,m. 
Additionally to the mechanical loading, the bottom surface of the specimen is loaded by a~water pressure of 0.3\,MPa. The water pressure at the upper surface is set to zero. The remaining, vertical surface are sealed, no flux is allowed perpendicularly to them.

The numerical simulations are conducted using material parameters shown in Table~\ref{tab:material_3pb}. The Biot coefficient was either 0, 0.5 or 1 to vary the intensity of coupling of the mechanical and transport problems. 
It is expected that the larger Biot coefficient results in a~lower peak mechanical loading force as the fluid pressure contributes more to the crack opening.

\begin{figure}[!b]
    \centering
   \begin{minipage}[t]{0.015\textwidth}
   \rotatebox{90}{\small\textbf{\emph{adaptive} m.}}
    \end{minipage}
    \begin{minipage}[t]{0.02\textwidth}
   \rotatebox{90}{\small peak load}
    \end{minipage}
        \begin{minipage}[t]{0.35\textwidth}
        \includegraphics[width=1\textwidth]{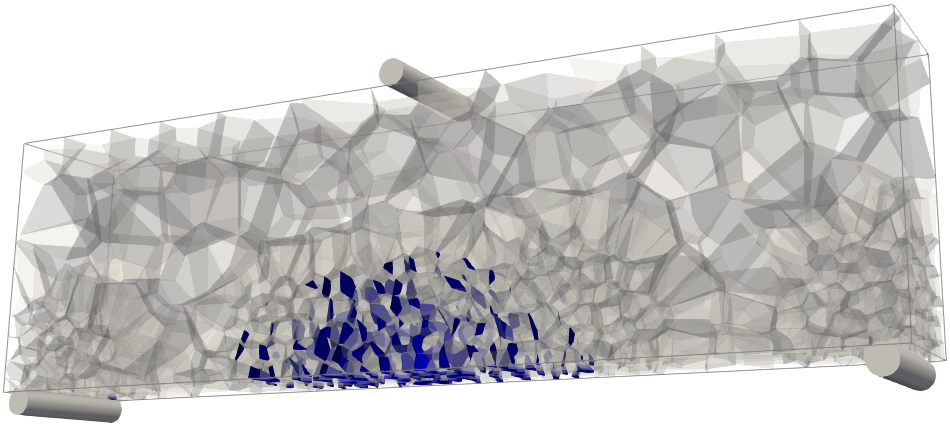}
        
        \vspace{-2.9cm}\:\:\:\fcolorbox{black}{black!5}{\small10966 DoF}
        
        \vspace{-1.5cm}
        \begin{center}
           \small Biot coefficient = 0
        \end{center}  
    \end{minipage}
     \begin{minipage}[t]{0.35\textwidth}
        \includegraphics[width=1\textwidth]{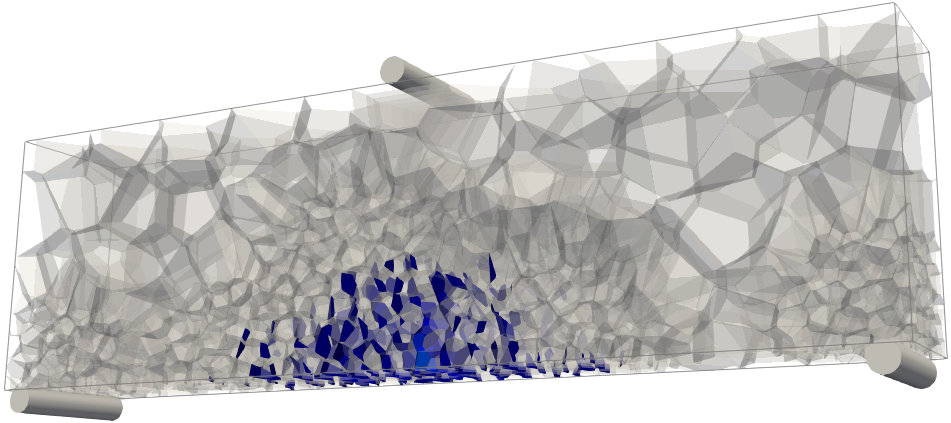}
        
        \vspace{-2.9cm}\:\:\:\fcolorbox{black}{black!5}{\small15456 DoF}
        
        \vspace{-1.5cm}
        \begin{center}
        \small    Biot coefficient = 1
        \end{center}  
    \end{minipage}
    
 \begin{minipage}[t]{0.015\textwidth}
   \rotatebox{90}{\small\textbf{\emph{adaptive} m.}}
    \end{minipage}
    \begin{minipage}[t]{0.02\textwidth}
   \rotatebox{90}{\small terminal state}
    \end{minipage}
     \begin{minipage}[t]{0.35\textwidth}
       \includegraphics[width=1\textwidth]{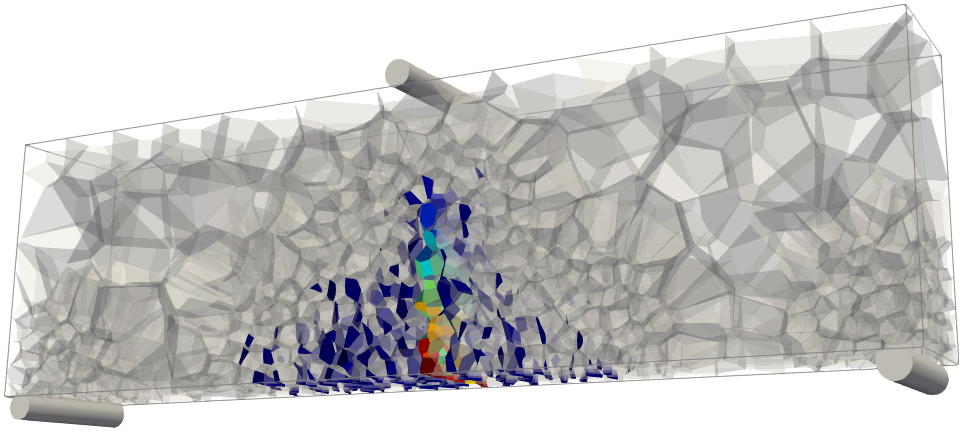}
       
       \vspace{-2.9cm}\:\:\:\fcolorbox{black}{black!5}{\small 13340 DoF}
    \end{minipage}
    \begin{minipage}[t]{0.35\textwidth}
        \includegraphics[width=1\textwidth]{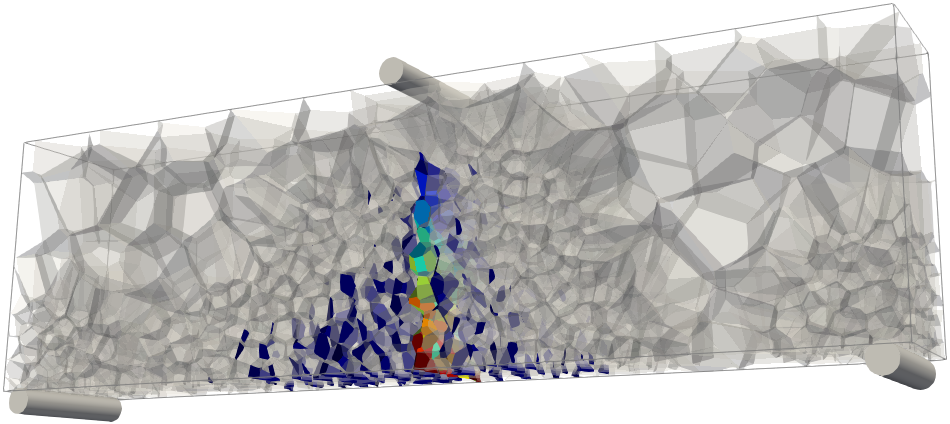}
        
       \vspace{-2.9cm}\:\:\:\fcolorbox{black}{black!5}{\small 16582 DoF}
    \end{minipage}
    
    \begin{minipage}[t]{0.015\textwidth}
   \rotatebox{90}{\small \textbf{\emph{fine} model}}
    \end{minipage}
    \begin{minipage}[t]{0.02\textwidth}
    \rotatebox{90}{\small terminal state}
    \end{minipage}
     \begin{minipage}[t]{0.35\textwidth}
        \includegraphics[width=1\textwidth]{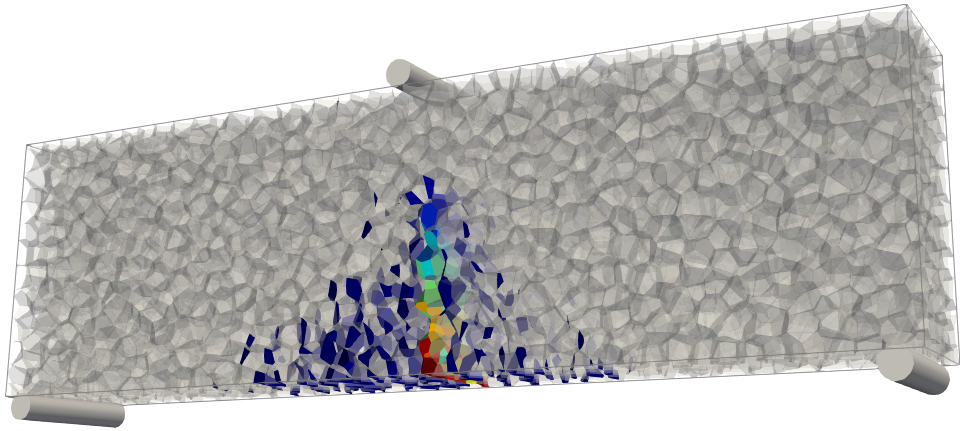}
        
        \vspace{-2.9cm}\:\:\:\fcolorbox{black}{black!5}{\small 50890 DoF}
    \end{minipage}
    \begin{minipage}[t]{0.35\textwidth}
        \includegraphics[width=1\textwidth]{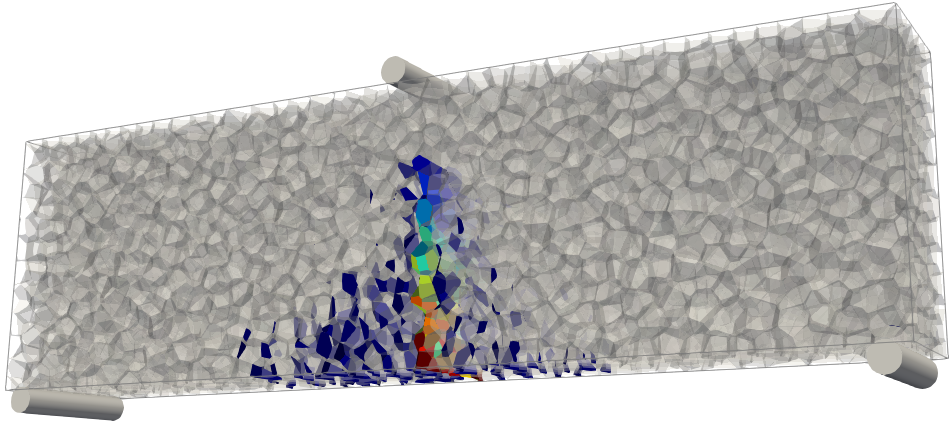}
        
        \vspace{-2.9cm}\:\:\:\fcolorbox{black}{black!5}{\small 50890 DoF}

    \end{minipage}
       \caption{
           Three point bending combined with fluid pressure: crack patterns obtained using the fine ($l_{\min}=20$\,mm) and
     adaptive models ($l_{\min}=64\rightarrow 20$\,mm). }
    \label{fig:3pb_cracks}
\end{figure}

\begin{table}[htbp]
\centering
\begin{tabular}{llll}

\multicolumn{4}{c}{mechanical properties}    \\\hline
normal modulus & $E_0$ &  60$\times10^9$ & Pa \\
tang. to norm. stiffness ratio & $\alpha$ & 0.29 & [-]  \\
tensile strength                     & $f_{\mathrm{t}}$ & 2.2$\times10^6$ & Pa  \\
fracture energy in tension                     & $G_{\mathrm{t}}$ & 35 & N/m \\      \hline \\
\multicolumn{4}{c}{transport properties} \\\hline
permeability  & $\kappa$ & $5\times10^{-18}$ & $\mathrm{m}^2$ \\
tortuosity          &     $\xi$  & 1  & [-]               \\      
viscosity           &  $\mu$    & $8.9\times10^{-4}$   & $\mathrm{Pa}\cdot\mathrm{s}$ \\
density           &  $\rho$    & 1000   & $\mathrm{kg}/\mathrm{m}^3$    \\\hline

\end{tabular}
 \caption{Material properties used in numerical simulations of three point bending combined with pressure load.}
    \label{tab:material_3pb}
\end{table}

\subsubsection{Numerical results}

\begin{figure}[!t]
    \centering
    \includegraphics[width=18cm]{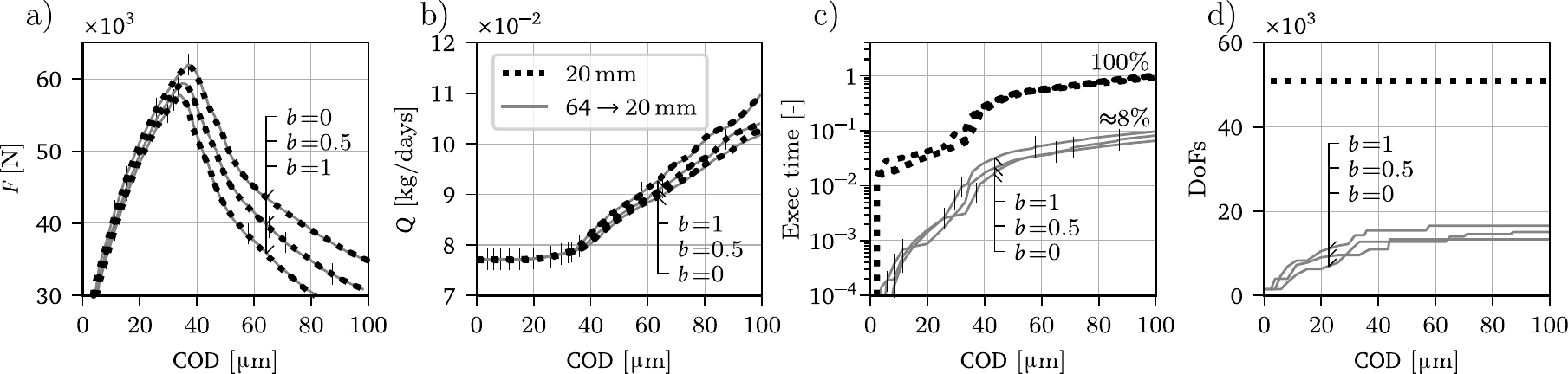}
    \caption{
    Three point bending combined with fluid pressure -- deterministic comparison of \emph{fine} ($l_{\min}=20$\,mm) and \emph{adaptive} ($l_{\min}=64\rightarrow20$\,mm) models{\color{black}:
    a) the loading force,
    b) the mass flux of water,
    c) the execution time and
    d) the number of degrees of freedom.
    }
}
    \label{fig:3pbsingle}
\end{figure}

First, a~deterministic comparison of response of the \emph{fine} ($l_{\min}=20$mm) and \emph{adaptive} ($l_{\min}=64\rightarrow20$\,mm) models is investigated. The discrete mesoscale model response is partially random due to random internal structure. The \emph{physical} discretization in the \emph{adaptive} model is, for the purpose of the deterministic comparison, given by identical generator nodes as in the \emph{fine} model to suppress the differences due to the random model heterogeneity. Fig~\ref{fig:3pbsingle}a shows dependency of the loading mechanical force, $F$, on the crack opening displacement (COD) for three different values of the Biot coefficient. The water mass flow between the bottom and top surface of the specimen is presented in Fig~\ref{fig:3pbsingle}b. The relative execution times are compared in Fig~\ref{fig:3pbsingle}c. The number of degrees of freedom of individual models is illustrated in Fig~\ref{fig:3pbsingle}d. The refinement events are highlighted by the short vertical lines across the plots of the results from \emph{adaptive} simulations.

The responses as seen in Figs~\ref{fig:3pbsingle}a,b are virtually identical. Note that the curves do not start at zero loading force. This is because the constant pressure loading, imposed before the first regular simulation step while the mechanical supports are fixed in the vertical direction, immediately induces mechanical reaction forces. Note that the maximum loading force is reached for Biot coefficient 0 when mechanics is completely independent on mass transport. The lowest peak load is observed for $b=1$ when the transfer of the fluid pressure inside the cracks into the solid phase is maximized. The crack patterns resulting from these simulations are shown in Fig.~\ref{fig:3pb_cracks}. The coloring illustrates the values of crack opening between individual grains. These cracks obtained by the \emph{fine} and \emph{adaptive} models look identical. One can also see the coarse internal structure of the \emph{adaptive} models in the elastic regions.

The \emph{adaptive} model solution was executed within $\approx8\,\%$ time compared to the \emph{fine} model ($\approx  12.5\times$ speed-up factor). The computational time savings are much larger in the earlier stages of the simulation.

\begin{figure}[!t]
    \centering
    \includegraphics[width=18cm]{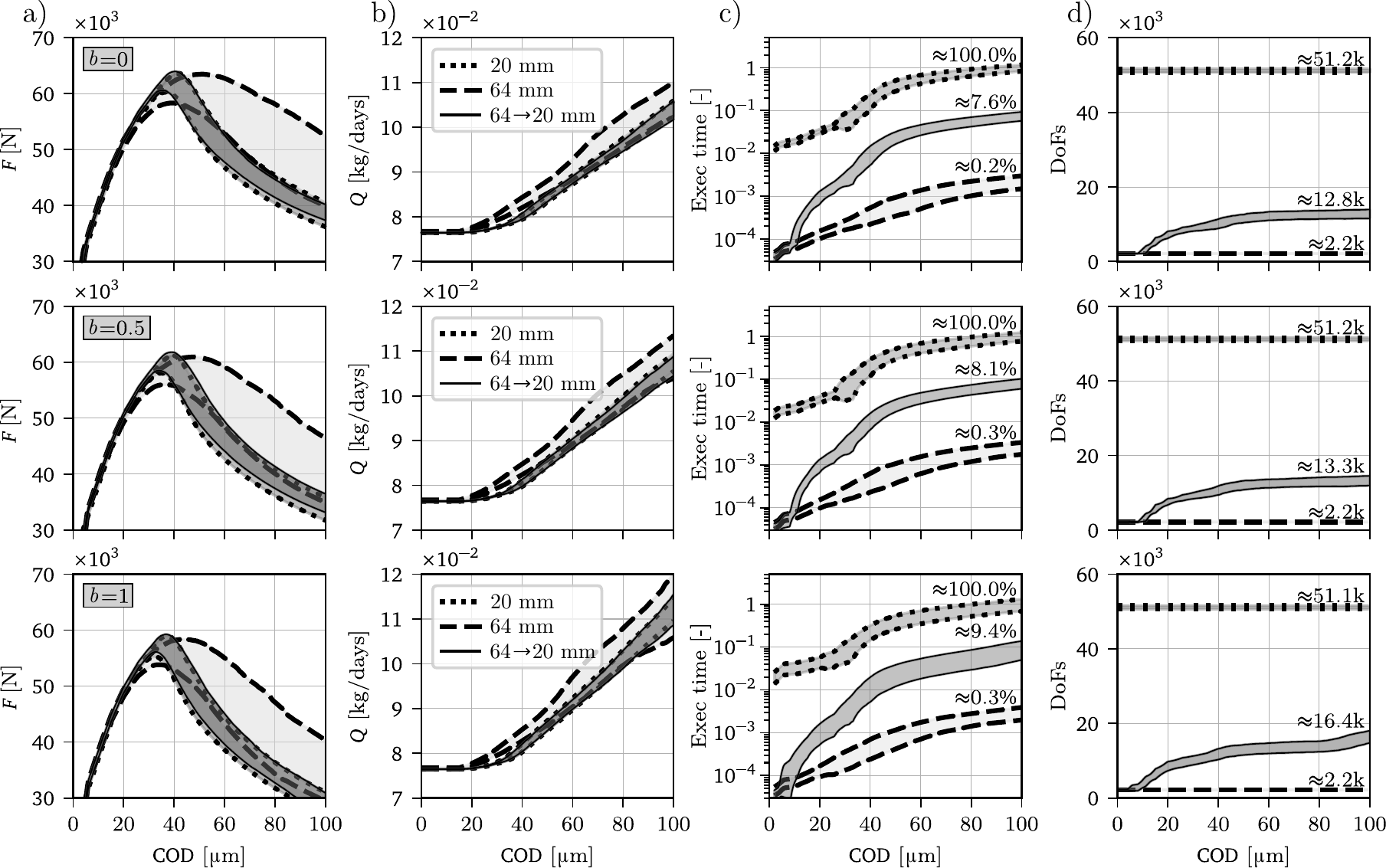}
    \caption{
    Three point bending combined with fluid pressure:
    a) the loading force,
    b) the mass flux of water,
    c) the execution time and
    d) the number of degrees of freedom.
}
    \label{fig:3pb_biots}
\end{figure}

Finally, a~statistical comparison between a~batch of 100 \emph{fine}, \emph{coarse} and \emph{adaptive} models was conducted. Unlike the deterministic comparison between two particular models structures, the adaptive refinement generated a~random fine structure on the fly in this case. Therefore a~precise match between fine and adaptive results is to be obtained only by conducting a~large number of realizations.

In Figs.~\ref{fig:3pb_biots}, the results are provided in a~similar fashion as before with each of rows displaying results for one Biot coefficient (0, 0.5, and 1, respectively). Instead of individual curves for each simulation, we present bands around mean response curve with thickness of one standard deviation at each side.
The \emph{adaptive} simulations provide results comparable to the \emph{fine} model while consuming $\approx8\,\%$ of the original execution time and by lowering the number of degrees of freedom to $\approx27\,\%$. The \emph{coarse} models with constant discretization $l_{\min}=64\,$mm is even faster, but there are large errors apparent especially in comparison of mechanical loading forces, but also in mass fluxes.

\subsection{Concrete spalling due to reinforcement bar corrosion}

\begin{figure}[!b]
    \centering
    \includegraphics[width=18cm]{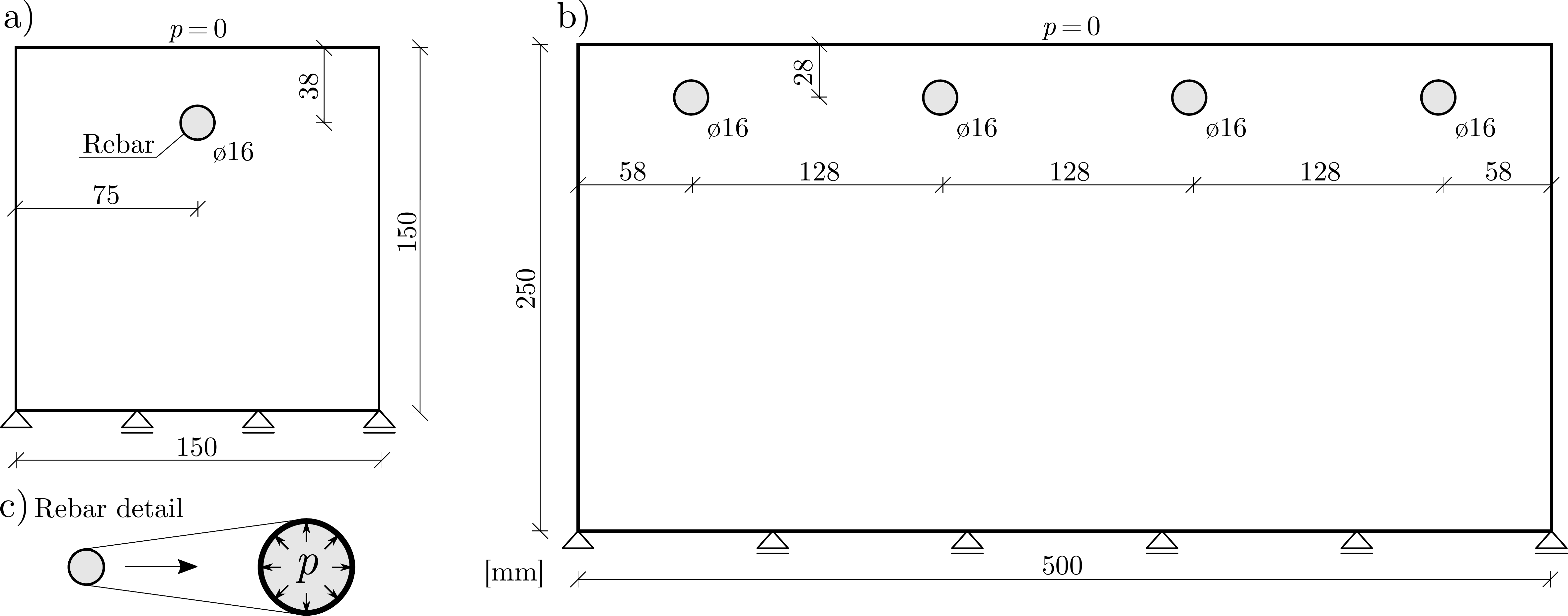}
    \caption{Corrosion induced cracking of concrete: a) single rebar model~\citep{andrade1993cover}, b) four rebars model according to Ref.~\citep{FahWhe-17}. }
    \label{fig:rebar}
\end{figure}

Concrete spalling due to the transformation of steel reinforcing bars into a~liquid corrosion products is a~coupled mass transport-mechanical problem. The corrosion products typically occupy a~volume greater than the original sound steel (2--6$\times$ according to Ref.~\citep{broomfield2003corrosion}).  Pressure gradient originating from volume expansion of liquid rust generates cracks in concrete and drives the transport of corrosion products into the surrounding concrete. The diffusion of fluid corrosion products through concrete significantly delays the beginning of corrosion induced cracking of concrete. Studies have shown \citep{wong2010penetration, michel2014penetration} that corrosion products are further transported into the more distant pores and cracks and may  additionally contribute to growth of existing cracks. 
{\color{black}
For further reading on the experimental and numerical research in the topic of steel corrosion, we redirect the reader to appropriate references such as  \cite{ahmad2003reinforcement,cabrera1996deterioration,castel2000mechanical}.
For measurements and diagnostics of the actual corroded profiles using acoustic emission, X-ray and 3D scanning, see, e.g., Refs.~\cite{sun2018evaluation,zhou2019experimental,tang2022investigation}.
%
%
The example of concrete spalling  due to reinforcement bar corrosion is included primarily to demonstrate possible application of the developed adaptive discretization for  coupled discrete model of mechanics and transport in concrete.
}

The physical experiment as conducted in Ref.~\cite{andrade1993cover} is simulated by a~2D discrete model, see Fig.~\ref{fig:rebar}. The same experiment was simulated by~\citet{FahWhe-17} along with additional geometry sketched in Fig.~\ref{fig:rebar}b. We tested the adaptive refinements strategy on both of these geometries. The first specimen is a~concrete block with a~cross-section of 150$\times$150\,mm$^2$; the second specimen is a~concrete block with a~cross-section of 500$\times$ 250\,mm$^2$. The steel rebars in both versions have diameter 16\,mm and the out-of-plane thickness is 1\,m. 
The 2D model with \emph{physical} discretization will still be called a~mesoscale model, but such simplification is very questionable. Many 3D effects of heterogeneity are missing. 
{\color{black}
Indeed, the simplification to a 2D problem neglects the heterogeneity across the third dimension and possible toughening mechanisms are omitted. 
The crack paths simulated in 2D are also missing the tortuousity from the third dimension, which affects the energy dissipation and transport properties. 
Consequently, with the same set of material parameters, a larger Poisson's ratio and elastic modulus but lower fracture energy and material strength shall be expected at the macroscale in two dimensions when compared to a full 3D model. 
Therefore, material parameters (or even character of the constitutive routines) of 2D models shall be phenomenologically modified from 3D models to account for the missing 3D discretization effect.
}
It would be more appropriate to understand the mesoscale character of the model only as a~way how to link size of the discrete units to the real material. 

To study the influence of mechanical-mass transport coupling, all simulations were conducted for values of the Biot coefficient of 0, 0.5, and 1, respectively.
The corrosion products are treated as an~incompressible fluid with time independent properties \cite{michel2014penetration}. Concrete is assumed to be fully saturated and Biot coefficient varies. The corrosion process is uniform along the steel/concrete interface along its circumference and also along the rebar length; it should however be noted that the overall model response may be strongly affected by the spatial distribution of corrosion~\citep{KunJir-21}.

Loading by an~increasing pressure of corrosion products takes place at the interface between the steel reinforcement bars and concrete, see Fig.~\ref{fig:rebar}c. The interface contains both transport and mechanical nodes. The pressure DoFs at the transport nodes are simply dictated by the prescribed external pressure, while the mechanical nodes are loaded by radial forces proportional to the area associated with the nodes and the prescribed pressure. The overall simulation, or more specifically the prescribed external pressure, is however controlled indirectly by an increase of the volume of the corrosion products in time.
More specifically, the prescribed variable is the change of the radius of the interface,  $x_{\mathrm{cor}}$, measured as an~average calculated from two perpendicular virtual displacement gauges across the interface circle.

{\color{black}
 The corrosion process is accelerated in laboratory experiments by applying electrical current to the steel reinforcement bars.
 It is possible to link the numerical experiment to the course of the physical experiment, as explained in detail in Ref.~\citep[Sec. 2]{FahWhe-17}:
the loss of steel layer, $dx_{\mathrm{cor}}$, can be related to the real-time increment in the physical experiment such as:
\begin{align}
    dx_{\mathrm{cor}}=0.0315\,i_{\mathrm{cor}}
\end{align}
where 0.0315 is a unit conversion factor from $\mu\mathrm{A}/\mathrm{cm}^2$ to $\mu\mathrm{m}$/days and $i_{\mathrm{cor}}$ is the current density (corrosion rate).
It is assumed that all the current is spent on steel dissolution without losses. No other contributions to the corrosion process are accounted for other than the induced corrosion.
Using this link, the loss of steel layer, $dx_{\mathrm{cor}}$, can be related to the layer of corrosion products by the expansion factor $\alpha_e$. The volume balance for the time change $dt$ reads:
\begin{align}
    \alpha_e dx_{\mathrm{cor}} =  dx_{\mathrm{cor}} + du_{\mathrm{cor}} + qdt
\end{align}
where $du_{\mathrm{cor}}$ is the displacement change at the interface between steel and concrete and $q$ is the flux of corrosion products transported into the surrounding concrete.}

The material parameters for both mechanical and transport model were set according to Ref.~\citep{FahWhe-17} and are listed in Tab.~\ref{tab:material2}. The expansion factor used in conversion of sound steel into corrosion products can be, according to Ref.~\citep{FahWhe-17}, taken as $\alpha_e=2$. Therefore, the density of the fluid corrosion products was assumed to be $\rho=3925\,\mathrm{kg/m^3}$, half of the typical sound steel density.

\begin{table}[tb!]
\centering
\begin{tabular}{llll}

\multicolumn{4}{c}{mechanical properties}    \\\hline
normal modulus & $E_0$ &  37$\times10^9$ & Pa \\
tang. to norm. stiffness ratio & $\alpha$ & 1 & [-]  \\
tensile strength                     & $f_{\mathrm{t}}$ & 3.2$\times10^6$ & Pa  \\
fracture energy in tension                     & $G_{\mathrm{t}}$ & 143 & N/m \\      \hline \\

\multicolumn{4}{c}{transport properties} \\\hline
permeability  & $\kappa$ & $1\times10^{-16}$ & $\mathrm{m}^2$ \\
tortuosity          &     $\xi$  & 0.001  & [-]               \\      
viscosity           &  $\mu$    & $1.9\times10^{4}$   & $\mathrm{Pa}\cdot\mathrm{s}$ \\
density           &  $\rho$    & 3925   & $\mathrm{kg}/\mathrm{m}^3$    \\\hline
\end{tabular}
 \caption{Material properties used in numerical simulations of corrosion spalling.}
    \label{tab:material2}
\end{table}

\begin{figure}[!b]
    \centering
    \includegraphics[width=18cm]{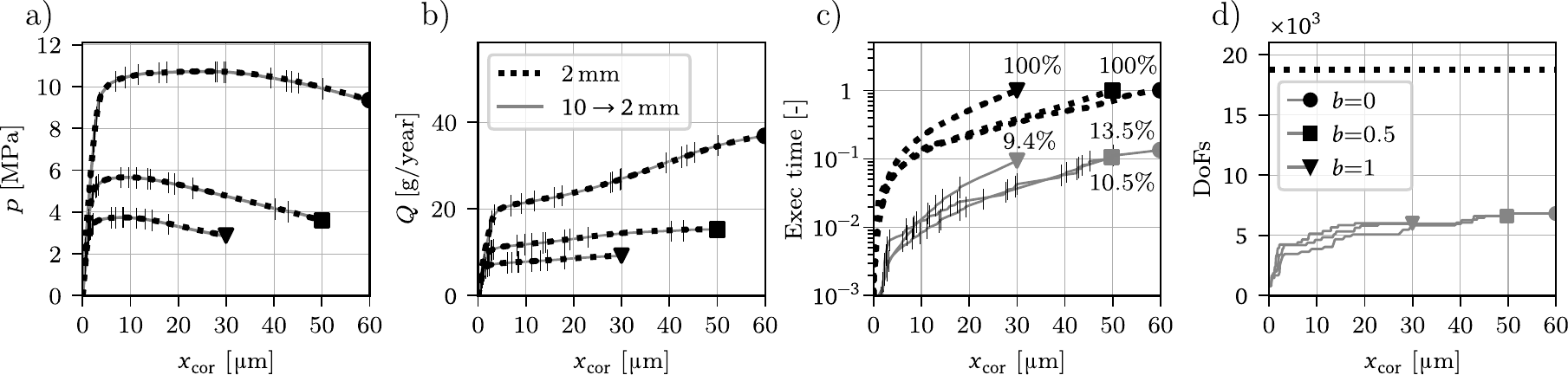}
    \caption{
    Corrosion induced cracking of a~single rebar model:
    a) the average fluid pressure along the steel/concrete interface,
    b) the total flux of fluid corrosion products through the steel/concrete interface,
    c) the execution time of the simulation,
    d) the number of degrees of freedom.
}
    \label{fig:singlerebarsingle}
\end{figure}

\begin{figure}[htbp]
    {\centering
    \hspace{3mm}
    \begin{minipage}[t]{0.22\textwidth}\centering
        \smaller Biot coefficient = 0
    \end{minipage}
    \begin{minipage}[t]{0.22\textwidth}\centering
        \smaller Biot coefficient = 1
    \end{minipage}
    
      \begin{minipage}[t]{0.015\textwidth}
    \rotatebox{90}{\smaller \quad\quad\quad\: \textbf{\emph{full} model} }
    \end{minipage}
    \begin{minipage}[t]{0.22\textwidth}
        \includegraphics[width=1\textwidth]{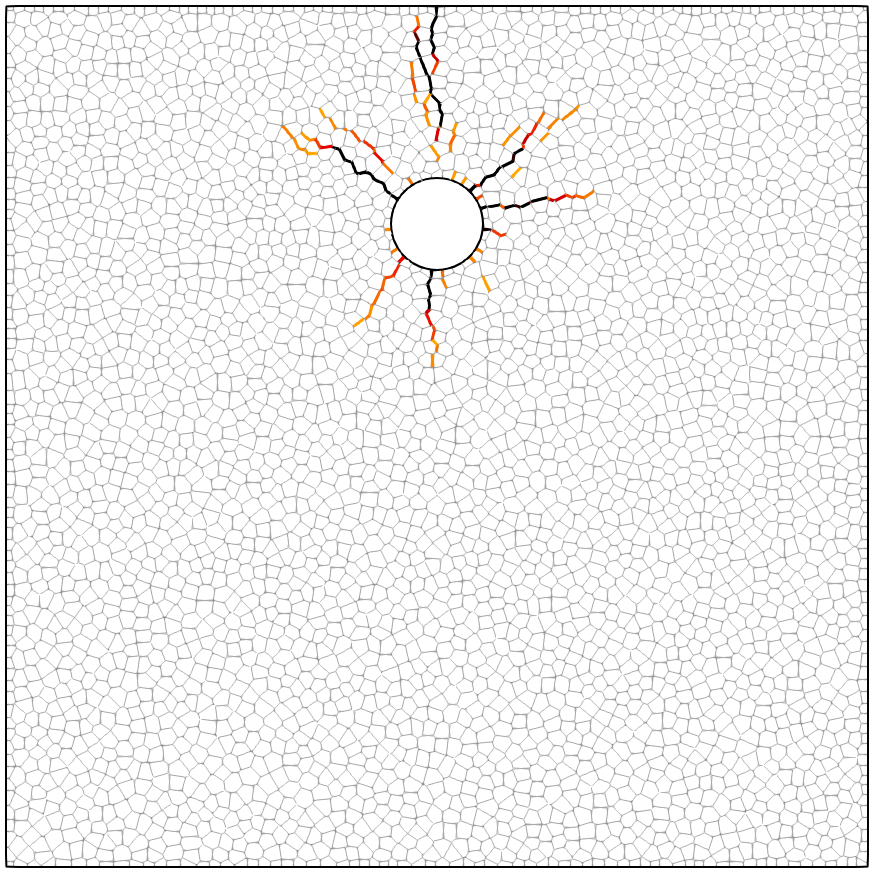}
    \end{minipage}
    \begin{minipage}[t]{0.22\textwidth}
        \includegraphics[width=1\textwidth]{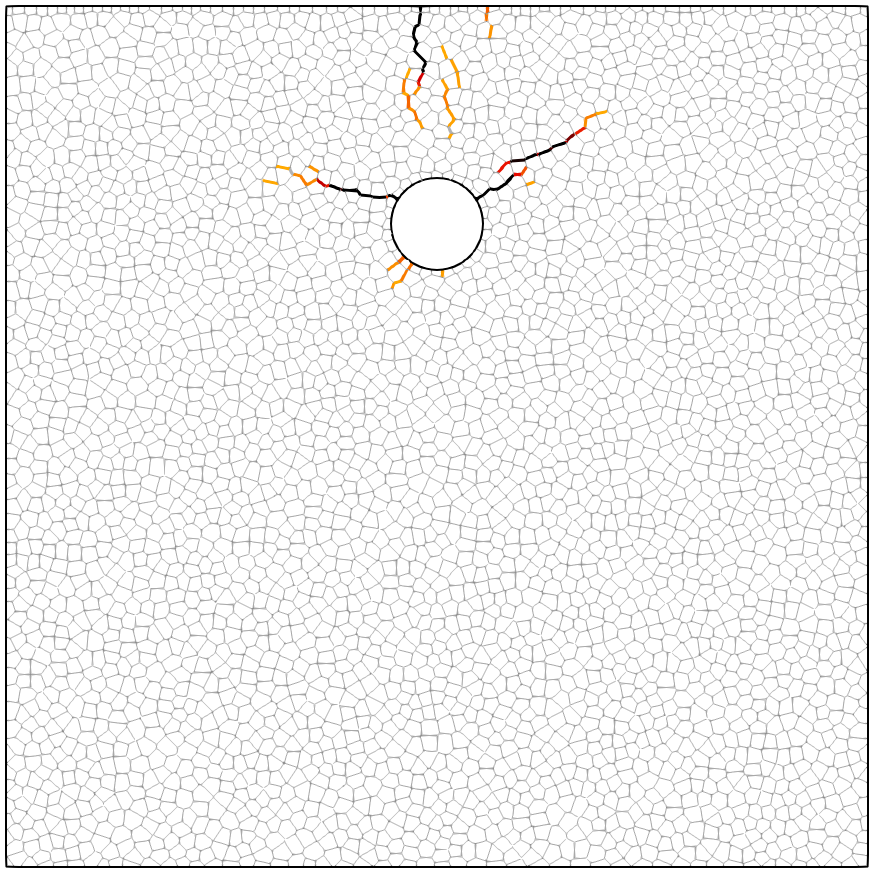}
    \end{minipage}

    \vspace{-0.8cm}
    \begin{minipage}[t]{0.22\textwidth}
       \smaller \centering \quad\: \fcolorbox{black}{black!5}{\smaller 18793 DoF}
    \end{minipage}
    \begin{minipage}[t]{0.22\textwidth}
      \smaller \centering \quad\: \fcolorbox{black}{black!5}{\smaller 18793 DoF}
    \end{minipage}
    
   \centering
   \vspace{0.3cm}
   \begin{minipage}[t]{0.015\textwidth}
    \rotatebox{90}{\smaller \quad\quad \textbf{\emph{adaptive} model} }
    \end{minipage}
     \begin{minipage}[t]{0.22\textwidth}
        \includegraphics[width=1\textwidth]{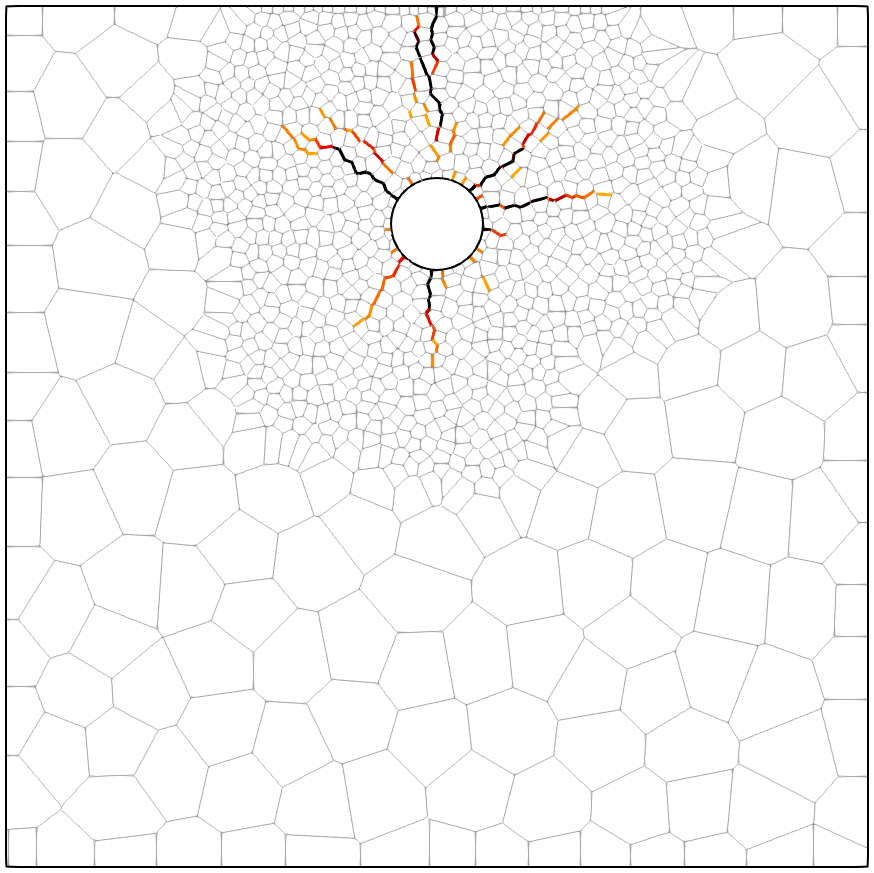}
    \end{minipage}\hspace{0.000\textwidth}
    \begin{minipage}[t]{0.22\textwidth}
        \includegraphics[width=1\textwidth]{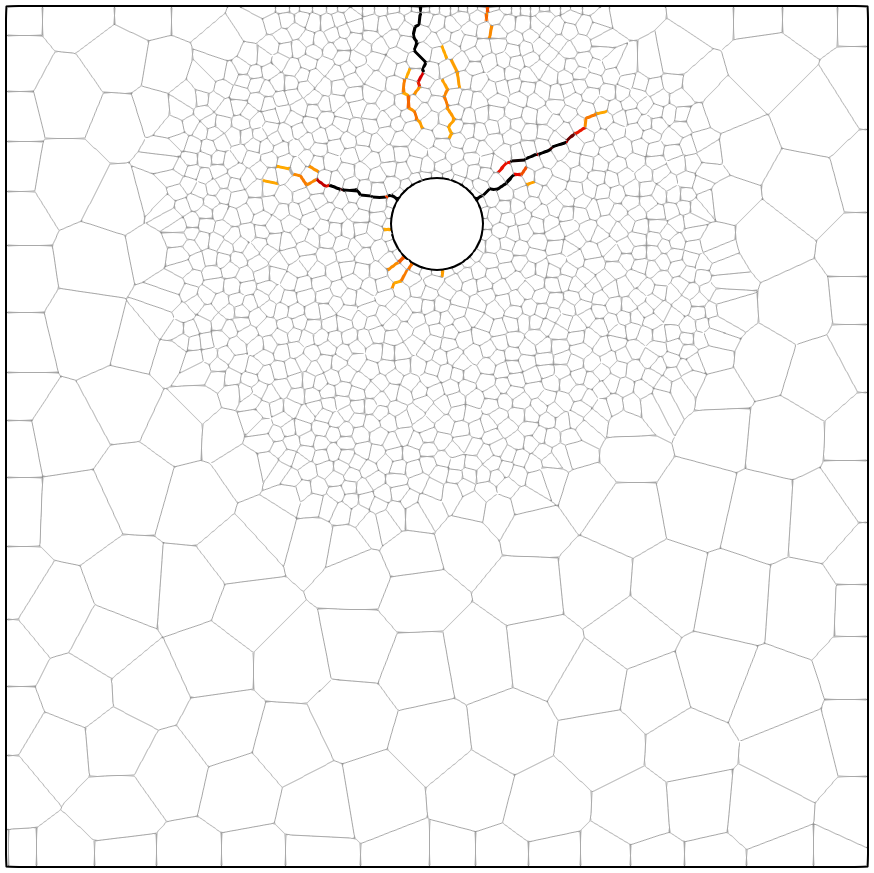}
    \end{minipage}

    \vspace{-0.8cm}
    \begin{minipage}[t]{0.24\textwidth}
     \smaller \centering \quad\quad\:\: \fcolorbox{black}{black!5}{\smaller 5078 DoF}
    \end{minipage}\hspace{0.004\textwidth}
    \begin{minipage}[t]{0.24\textwidth}
     \smaller \centering \fcolorbox{black}{black!5}{\smaller 5142 DoF}
    
    \vspace{0.3cm}  
     \hspace{0.05cm}     \includegraphics[width=3.6cm]{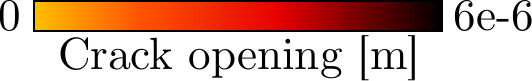}
    \end{minipage}

    }
    
    \caption{  Crack patterns at the peak fluid pressure obtained using \emph{fine} ($l_{\min}=2$\,mm) and
     \emph{adaptive} model ($l_{\min}=10\rightarrow2$\,mm). }
    \label{fig:singlerebarcracks}
\end{figure}

\begin{figure}[htbp]
    \centering
    \includegraphics[width=18cm]{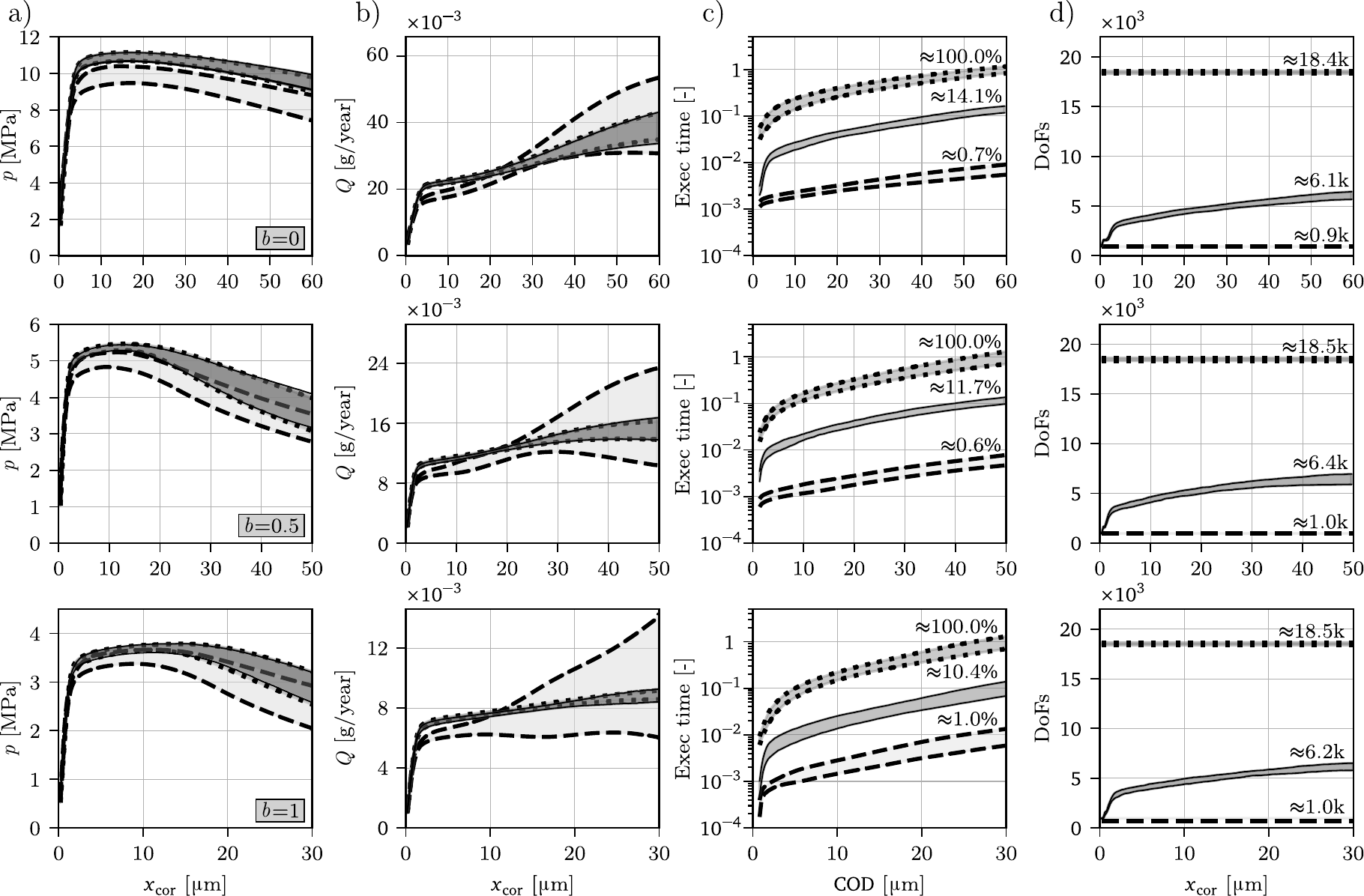}
    
       \caption{  Statistical response of 100 discrete coupled models:
    a) the average fluid pressure along the steel/concrete interface,
    b) the total flux of fluid corrosion products through the steel/concrete interface,
    c) the execution time of the simulation,
    d) the number of degrees of freedom.}
    \label{fig:singlerebarbatch}
\end{figure}

\subsubsection{Numerical results - single rebar model}
The results for the model with a~single rebar (Fig.~\ref{fig:rebar}a) are organized in a~very similar fashion for the three point bent beam. 
Fig.~\ref{fig:singlerebarsingle} presents a~deterministic comparison of responses of the \emph{fine} ($l_{\min}=2$mm) and the \emph{adaptive} ($l_{\min}=10\rightarrow2$mm) model. The refinement in the \emph{adaptive} model samples the same generator points as are present in the \emph{fine} model. The average fluid pressure along the steel/concrete interface in dependence on the loss of sound steel radius, $x_{\mathrm{cor}}$, shows in Fig.~\ref{fig:singlerebarsingle}a an~excellent match for all considered values of the Biot coefficient. The same correspondence is observed for the total flux of fluid corrosion products through the steel/concrete interface (Fig.~\ref{fig:singlerebarsingle}b). The execution times are compared in Fig.~\ref{fig:singlerebarsingle}c showing a~significant computational time reduction. Finally, the number of degrees of freedom of the models is shown by Fig.~\ref{fig:singlerebarsingle}d.
The adaptive refinement is again capable of capturing the mechanical and mass transport response provided by the \emph{fine} model while the speed-up factor delivered by the adaptive technique is $\approx10\times$.

The crack patterns resulting from these simulations are shown in Fig.~\ref{fig:singlerebarcracks}. The color scale correspond to openings of cracks between individual grains. The crack patterns obtained from the \emph{fine} and \emph{adaptive} model are practically identical.

The statistical comparison is presented next. The adaptive refinement is generated randomly on the fly so the comparison is reasonable only across a~large number of realizations. Similarly to the preceding example, a~batch of simulations of 100 \emph{adaptive}, \emph{fine} and \emph{coarse} models was executed to provide such statistical comparison. In Fig.~\ref{fig:singlerebarbatch}, the results are displayed for various values of Biot coefficient (0, 0.5 and 1). The graphs show bands around the mean curves with thickness of one standard deviation on each side. Again, the statistical comparison confirms that the \emph{adaptive} model delivers the same results as the \emph{fine} model; the \emph{coarse} model provides results with large error. The execution time needed by the \emph{adaptive} model is $\approx10-14\,\%$ of the execution time of the \emph{fine} model; the number of degrees of freedom is reduced to $\approx35\,\%$.

\clearpage
\subsubsection{Numerical results - four rebars model}

\begin{figure}[htbp]
    \centering
    \includegraphics[width=18cm]{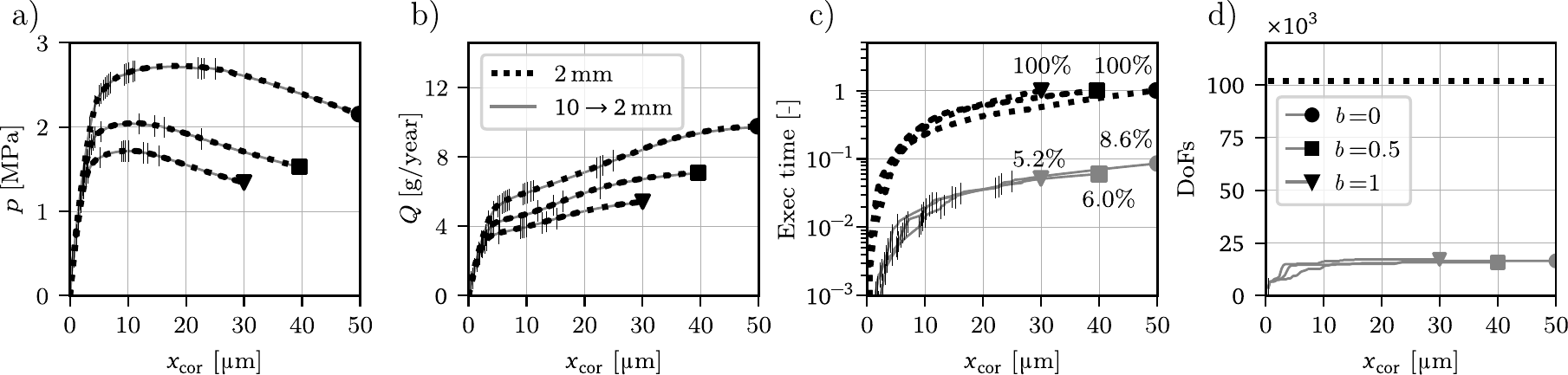}
    \caption{
    Corrosion induced cracking of a~four rebars model:
    a) the average fluid pressure along the steel/concrete interface,
    b) the total flux of fluid corrosion products through the steel/concrete interface,
    c) the execution time of the simulation,
    d) the number of degrees of freedom.
}
    \label{fig:fourrebarsingle}
\end{figure}

\begin{figure}[htbp]
    \centering
    
  \hspace{6mm}
     \begin{minipage}[t]{0.46\textwidth}
       \smaller \centering  Biot coefficient = 0
    \end{minipage}
    \begin{minipage}[t]{0.46\textwidth}
       \smaller \centering  Biot coefficient = 1
    \end{minipage}
    
     \begin{minipage}[t]{0.015\textwidth}
    \rotatebox{90}{\smaller \:\:\:\:\:\:\:\, \textbf{\emph{fine}} }
    \end{minipage}
    \begin{minipage}[t]{0.02\textwidth}
    \rotatebox{90}{\smaller \:\:\:\:\: \textbf{model} }
    \end{minipage}
     \begin{minipage}[t]{0.46\textwidth}
        \includegraphics[width=1\textwidth]{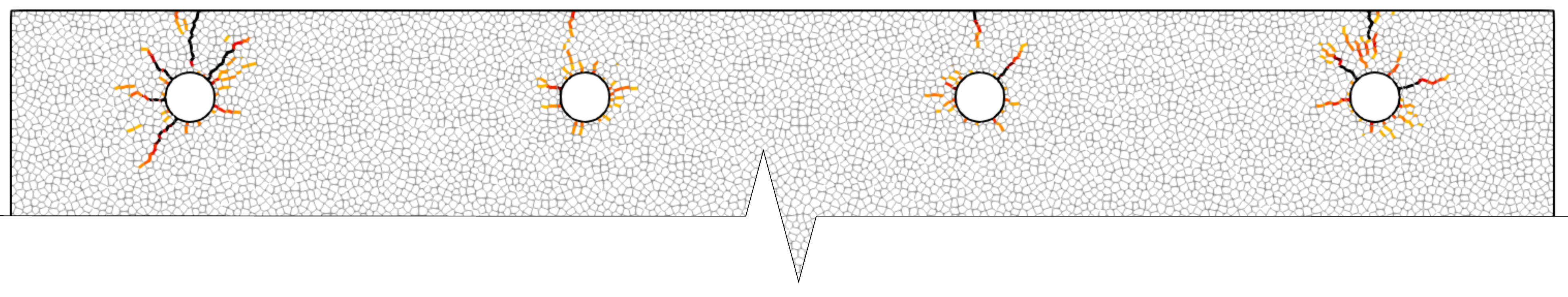}
        
        \vspace{-0.45cm}  \:\:\:\:\fcolorbox{black}{black!5}{\smaller 101943 DoF}
    \end{minipage}
    \begin{minipage}[t]{0.46\textwidth}
       \includegraphics[width=1\textwidth]{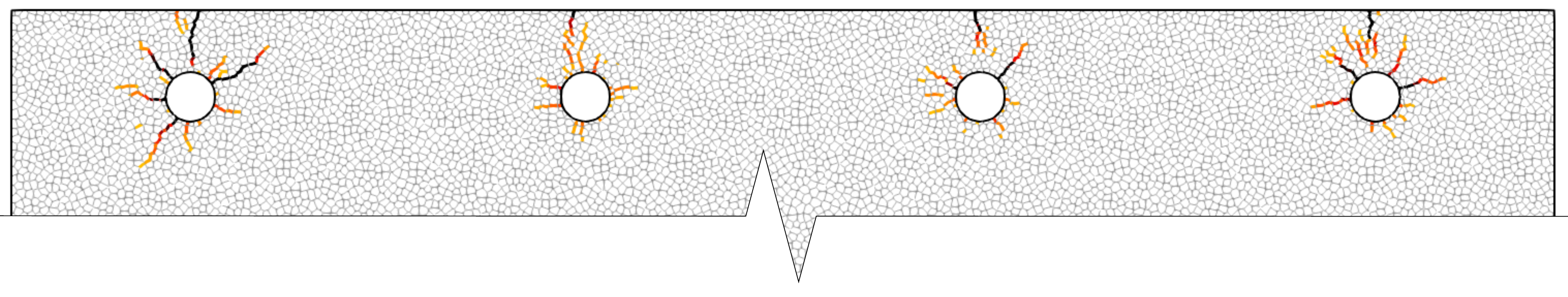}
       
        \vspace{-0.45cm}  \:\:\:\:\fcolorbox{black}{black!5}{\smaller 101943 DoF}
    \end{minipage}
    
  \begin{minipage}[t]{0.015\textwidth}
    \rotatebox{90}{\smaller \:\: \textbf{\emph{adaptive}} }
    \end{minipage}
    \begin{minipage}[t]{0.02\textwidth}
    \rotatebox{90}{\smaller \:\:\:\:\: \textbf{model} }
    \end{minipage}
     \begin{minipage}[t]{0.46\textwidth}
            \includegraphics[width=1\textwidth]{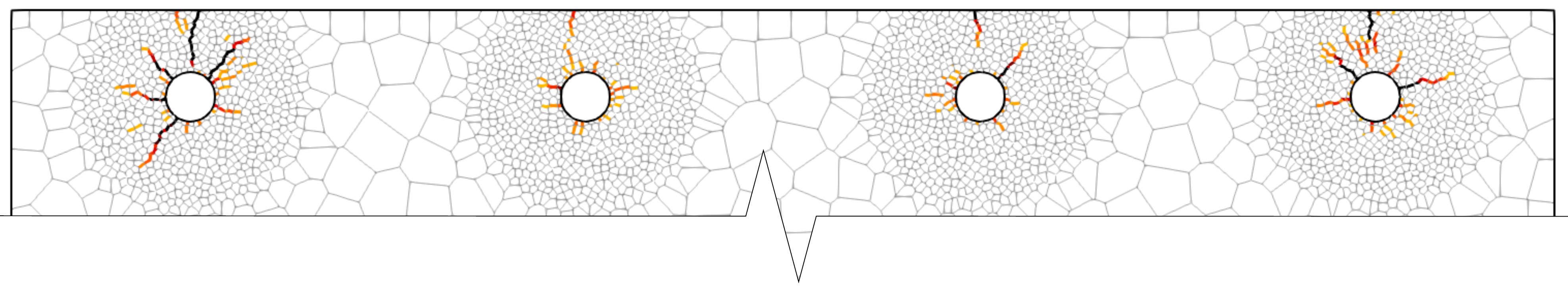}
            
             \vspace{-0.45cm}  \:\:\:\:\fcolorbox{black}{black!5}{\smaller 16503 DoF}
    \end{minipage}
    \begin{minipage}[t]{0.46\textwidth}
        \includegraphics[width=1\textwidth]{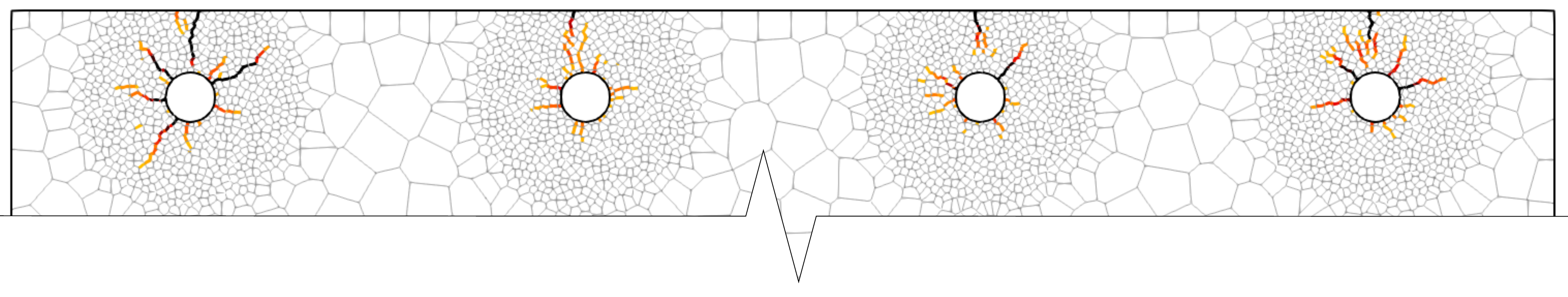}
        
         \vspace{-0.45cm}  \:\:\:\:\fcolorbox{black}{black!5}{\smaller 17204 DoF}
    \end{minipage}
    
    \vspace{-0.4cm}
    \begin{minipage}[t]{1\textwidth}
        \hspace{14.1cm}\includegraphics[width=3.6cm]{range_rebars.pdf}
    \end{minipage}

    \caption{  Crack patterns at the peak pressure obtained using the \emph{fine} ($l_{\min}=2$\,mm) and
     the \emph{adaptive} model ($l_{\min}=10\rightarrow2$\,mm). }
    \label{fig:fourrebarcracks}
\end{figure}

The studies of simulations of reinforcement bar corrosion process are completed by a~2D model with spatially larger domain that contains four reinforcing bars, Fig.~\ref{fig:rebar}b. Fig.~\ref{fig:fourrebarsingle} shows deterministic comparison of the \emph{fine} ($l_{\min}=2$\,mm) and the \emph{adaptive} ($l_{\min}=10\rightarrow2$\,mm) model, again the adaptive refinement used the generator points from the \emph{fine} model.
The average fluid pressure along the steel/concrete interface, the total flux of fluid corrosion products through the steel/concrete interface, the execution times of both simulations, and the number of degrees of freedom are presented in Fig.~\ref{fig:fourrebarsingle}a-d. The adaptive refinement solution is again well capable of capturing the mechanical and mass transport response yielded by the \emph{fine} model. The speed-up factor delivered by the \emph{adaptive} model at the end of the simulation is $\approx12-20\times$, consuming about $5-8\,\%$ of the execution time, depending on the value of the Biot coefficient. The number of degrees of freedom is reduced to $16\,\%$ when using the adaptive model as the damage and cracking is present only within a~small portion of the specimen. The crack patterns resulting from these simulations are shown in Fig.~\ref{fig:fourrebarcracks} with colors corresponding to crack openings in the mechanical discrete elements.

\clearpage
\section{Conclusions}

An  adaptive refinement of discretization is developed for steady state coupled discrete models of mechanics and mass transport in fully saturated medium. The coupling is provided by (i) the Biot's theory of poroelasticity and (ii) the effect of cracks on material permeability tensor.
\begin{itemize}
\item The mechanical model, the mass transport model and the Biot's coupling are independent in the elastic regime on the size of the discrete bodies, i.e., on the density of Voronoi generator points. The simulation therefore starts with a coarse, \emph{non-physical} discretization.

\item The Rankine criterion is used to detect locations where the stress state approaches inelastic regime. The tensorial stress is evaluated from tractions at the contacts of individual particles in the coarse discretization. When the maximum principal stress exceeds 70\,\% of material tensile strength, discretization in that region is refined. There is no transfer of history variables because only elastic part of the domain are replaced.

\item The refinement strategy is tested in (i) 3D on concrete beams loaded in three point bending combined with fluid pressure at the bottom face, and in (ii) 2D on simple models motivated by concrete fracturing due to corrosion of reinforcement bars. The Biot coefficient is set to 0, 0.5, or 1, respectively. The deterministic and statistical comparison shows an excellent match of results obtained with the \emph{fine} model with \emph{physical} discretization and the \emph{adaptive} model.

\item In all the cases, a~significant speed-up factor of $\approx8-20\times$ was obtained by reducing the number of degrees of freedom to~$\approx16-35\,\%$, depending on the proportion of domain experiencing inelastic behavior. 
\end{itemize}

In conclusion, it has been demonstrated that the adaptive refinement algorithm can be successfully and effectively applied to two-way coupled discrete mesoscale models of mechanics and mass transport.

\section*{Acknowledgement}
Financial support from the Czech Science Foundation under project no. 
22-06684K
is gratefully acknowledged.

\bibliographystyle{plainnat}
\bibliography{inner}

\end{document}